\documentclass[prb,aps,amssymb,onecolumn,superscriptaddress,notitlepage]{revtex4-1}
\usepackage{amsmath}
\usepackage{amssymb}
\usepackage{amsthm}
\usepackage{amsfonts}
\usepackage{mathrsfs}
\usepackage{listings}
\usepackage{lmodern}
\usepackage{enumerate}
\usepackage{latexsym}
\usepackage{rotating}
\usepackage{float}
\usepackage{psfrag}
\usepackage{bm}
\usepackage{graphicx}
\usepackage[caption=false]{subfig}
\usepackage{blkarray}
\usepackage{array}
\usepackage{color}
\usepackage[normalem]{ulem}
\usepackage{hyperref}

\newcolumntype{L}{>{$}l<{$}}

\newtheorem*{defn*}{Definition}

\newcommand{\half}{\frac{1}{2}}
\newcommand{\miniscule}{\fontsize{4}{12}} 
\begin{document}
\title{Graph Theory Data for \emph{Topological Quantum Chemistry}}
\author{M.~G. Vergniory}
\thanks{These authors contributed equally to the preparation of this work.}
\affiliation{Donostia International Physics Center, P. Manuel de Lardizabal 4, 20018 Donostia-San Sebasti\'{a}n, Spain}
\affiliation{Department of Applied Physics II, University of the Basque Country UPV/EHU, Apartado 644, 48080 Bilbao, Spain}
\affiliation{Max Planck Institute for Solid State Research, Heisenbergstr. 1,
70569 Stuttgart, Germany.}
\author{L. Elcoro}
\thanks{These authors contributed equally to the preparation of this work.}
\affiliation{Department of Condensed Matter Physics, University of the Basque Country UPV/EHU, Apartado 644, 48080 Bilbao, Spain}
\author{Zhijun Wang}
\affiliation{Department of Physics, Princeton University, Princeton, New Jersey 08544, USA}
\author{Jennifer Cano}
\affiliation{Princeton Center for Theoretical Science, Princeton University, Princeton, New Jersey 08544, USA}

\author{C. Felser}
\affiliation{Max Planck Institute for Chemical Physics of Solids, 01187 Dresden, Germany}
\author{M.~I.~Aroyo}
\affiliation{Department of Condensed Matter Physics, University of the Basque Country UPV/EHU, Apartado 644, 48080 Bilbao, Spain}
\author{B. Andrei Bernevig}
\affiliation{Department of Physics, Princeton University, Princeton, New Jersey 08544, USA}
\affiliation{Donostia International Physics Center, P. Manuel de Lardizabal 4, 20018 Donostia-San Sebasti\'{a}n, Spain}
\thanks{On sabbatical}
\affiliation{Laboratoire Pierre Aigrain, Ecole Normale Sup\'{e}rieure-PSL Research University, CNRS, Universit\'{e} Pierre et Marie Curie-Sorbonne Universit\'{e}s, Universit\'{e} Paris Diderot-Sorbonne Paris Cit\'{e}, 24 rue Lhomond, 75231 Paris Cedex 05, France}
\thanks{On sabbatical}
\affiliation{Sorbonne Universit\'{e}s, UPMC Univ Paris 06, UMR 7589, LPTHE, F-75005, Paris, France}
\thanks{On sabbatical}
\author{Barry Bradlyn}
\affiliation{Princeton Center for Theoretical Science, Princeton University, Princeton, New Jersey 08544, USA}
\date{\today}
\begin{abstract}
Topological phases of noninteracting particles are distinguished by global properties of their band structure and eigenfunctions in momentum space. On the other hand, group theory as conventionally applied to solid-state physics focuses only on properties which are local {(at high symmetry points, lines, and planes)} in the Brillouin zone. To bridge this gap, we {have previously [B. Bradlyn et al., \emph{Nature} {\bf 547}, 298--305 (2017)] mapped} the problem of constructing global band structures out of local data to a graph construction problem. {In this paper, we provide the explicit data and formulate the necessary algorithms to produce all topologically distinct graphs. Furthermore, we show how} to apply these algorithms to certain ``elementary'' band structures highlighted in the aforementioned reference, and so identified and tabulated all orbital types and lattices that can give rise to topologically disconnected band structures. {Finally, we show how to use the newly developed BANDREP program on the Bilbao Crystallographic Server to access the results of our computation.}
\end{abstract}
\maketitle
\section{Background \& Summary}

One of the most unexpected developments in condensed matter physics was the recent discovery of noninteracting topological insulators, in which the global dependence of Bloch wavefunctions on crystal momentum is topologically distinct from that in the atomic limit. While the original classification of such topological phases incorporated only on-site symmetries such as time-reversal, charge conjugation, and particle hole transformation\cite{Schnyder2008,KitaevClassify,ryu2010topological}, the set of distinct topological insulators protected by crystal symmetry is much richer\cite{Fu2011,Freed2013,Chiu2014,Shiozaki2014,Alexandradinata14,Hourglass,ArisCohomology,Shiozaki2017}. However, the conventional application of crystal symmetry to band theory is in the local study of degeneracies at isolated high-symmetry points in the Brillouin zone, through the $\mathbf{k}\cdot\mathbf{p}$ method\cite{Kittel87,bradley}. In this approach, one studies the subset of crystal symmetries that leave a particular, isolated $\mathbf{k}$-vector invariant. Electronic states at this momentum point transform under irreducible representations of the little group, and Hamiltonians in the neighborhood of this wavevector can be perturbatively expanded in terms of these representations. This local approach obfuscates any connection to band topology, which requires understanding how these different localized descriptions fit together to make a global band structure. While there has been some recent work towards addressing how symmetry constrains global band structures\cite{Kruthoff2016,Po2017}, a complete and constructive approach to the problem has not yet been presented.

On the other hand, we know from Heisenberg that global properties in momentum space map under the Fourier transform to local properties in position space. As such, one approach to topological band theory is to study the sets of energy bands that can arise from localized, symmetric orbitals in the atomic limit. This approach was introduced by the present authors in Ref.~\onlinecite{NaturePaper}, using the theory of induced ``band representations''. {We} argued that all sets of bands induced from symmetric, localized orbitals are topologically trivial by design; furthermore, by contraposition, any group of bands that does not arise as a band representation \emph{must} be topologically nontrivial. 

In order to determine whether or not a given set of bands transform as a band representation, however, requires knowing global information about the band structure. In particular topological phase transitions occur when isolated sets of topological bands disconnect from a band representation; cataloguing when these transitions can occur requires an understanding of the allowed connectivities of energy bands throughout the whole Brillouin zone. Thus, while the real-space approach makes the topology of Bloch bands manifest, it cannot be practically useful unless we directly address the issue of band connectivity in momentum space. 

In order to solve this problem, we {introduced and briefly discussed in Ref.~\onlinecite{NaturePaper} a mapping from} the problem of patching together the local, $\mathbf{k}\cdot\mathbf{p}$ bands to form a global band structure to a problem in graph theory. {As described in full theoretical detail in the subsequent Ref.~\onlinecite{GraphTheoryPaper}}, we map the little group representations appearing at each high-symmetry $\mathbf{k}$-vector to a node in a ``connectivity graph''. Edges in this graph are drawn in accordance with the group-theoretic ``compatibility relations'' between representations, as defined in Refs.~\onlinecite{bradley,kvec,grouptheory}. Given a set of irreducible representations at every $\mathbf{k}$-vector -- chosen for instance from an \emph{elementary band representation} (EBR) -- the problem of enumerating all the global ways in which these representations can be connected is equivalent to the problem of enumerating all distinct connectivity graphs. 

In the present work, we outline the algorithms we have developed both to construct these connectivity graphs, and to enumerate all the elementary band representations that allow for topologically disconnected band structures, both with and without spin-orbit coupling and time-reversal symmetry. In Secs.~\ref{sec:paths}--\ref{subsec:nonsymmorphic} we present algorithms for finding the minimal set of paths through the Brillouin zone which fully determine the connectivity of a band structure. Next, after reviewing the formal aspects of the mapping of band connectivity to graph theory, we present two algorithms for constructing and identifying disconnected connectivity graphs. The first approach, given in Secs.~\ref{sec:maiagraph} and \ref{sec:maialaplacian}, involves a direct implementation of the group-theoretic constraints on the connectivity graph, and the application of spectral graph theory\cite{GraphThy} to identify disconnected subgraphs (bands). The second approach, discussed in Sec.~\ref{sec:luisgraph}, builds disconnected subgraphs directly, growing outward from the $\mathbf{k}$-vector with the fewest nodes, in the spirit of Prim's algorithm\cite{prim1957shortest}. Throughout the discussion, we use the non-symmorphic space group $P4/ncc$ ($130$) as an illustrative example. We have applied these algorithms to tabulate the allowed connectivities of all band representations. We have made {the entirety of} this data available in the form of end-user programs, whose output is described in Sec.~\ref{sec:filedesc}. In Sec.~\ref{sec:valid} we show how to apply our data to the physically relevant example of graphene, the prototypical (symmorphic) topological insulator; this also serves as a consistency check on our algorithms. Finally, in Sec.~\ref{sec:notes}, we show how to access and utilize our data.

\section{Methods} \label{sec:methods}

\subsection{Determination of minimal $\mathbf{k}$-vectors and paths}\label{sec:paths}

To begin to approach the problem of computing all distinct connectivity graphs for each space group, we must first enumerate a minimal set of paths through the Brillouin zone that are needed for the computation, as described in Ref.~\onlinecite{GraphTheoryPaper}. To that end, 
we have developed an algorithm which identifies {a priviledged set of \emph{maximal} $\mathbf{k}$-vectors (defined below)} for all the 230 space groups, and the minimal set of non-redundant connections between them needed to construct a connectivity graph. The details of the theory and definitions used in the algorithm are described in Ref.~\onlinecite{grouptheory}, although we summarize the essentials here. 

Given a space group {$G$} and a $\mathbf{k}$-vector in reciprocal space, the symmetry operations $R$ of the point group that keep $\mathbf{k}$ invariant modulo a reciprocal lattice translation -- i.~e.~, those that fulfill the relation 
\begin{equation}
\mathbf{k}R=\mathbf{k}+\mathbf{K},\label{eq:cogroup}
\end{equation}
with $\mathbf{K}$ any vector of the reciprocal lattice -- belong to the little co-group $\bar{G}_\mathbf{k}$ of $\mathbf{k}$. {Note that to be consistent with the crystallographic conventions of Ref.~\onlinecite{grouptheory}, we define the little co-group as acting on $\mathbf{k}$-vectors from the right.} The little co-group of a $\mathbf{k}$-vector is isomorphic to one of the $32$ crystallographic point groups. 
For each $\mathbf{k}$-vector $\mathbf{k}_i$, we consider the {(closure of\footnote{It is important to take the closure of the manifolds defined in this way, in order to ensure that the high symmetry points at the endpoints of lines are defined to be contained in the lines, etc.})} largest continuous submanifold of reciprocal space which contains $\mathbf{k}_i$ such that the little co-group of every point in the manifold is isomorphic to $\bar{G}_{\mathbf{k}_i}$.
{We refer to such a submanifold, either a point, line, plane, or volume, of $\mathbf{k}$-vectors as a \emph{$\mathbf{k}$-manifold}. Each $\mathbf{k}$-manifold is identified by a letter}, and is specified by coordinate triplets depending on 0,1,2 and 3 free parameters, respectively. {We emphasize the (rather tautological) fact that picking specific values for these free parameters yields a $\mathbf{k}$-vector contained in the $\mathbf{k}$-manifold.} We say that two {$\mathbf{k}$-manifolds $A=\{\mathbf{k}_1(\mathbf{u}_1)\}$ and $B=\{\mathbf{k}_2(\mathbf{u}_2)\}$} are \emph{connected} if, {for some specific values of the free parameters $\mathbf{u}_1$ and $\mathbf{u}_2$, we have
\begin{equation}
\label{eq:connected_ks}
\mathbf{k}_1(\mathbf{u}_1)=\mathbf{k}_2(\mathbf{u}_2)+\mathbf{K}
\end{equation}
for some vector $\mathbf{K}$ of the reciprocal lattice.}

The star of a vector $\mathbf{k}$, {denoted $*\mathbf{k}$} is the set of vectors $\{\mathbf{k}R\}$ for all $R$ in the point group of the space group, which are \emph{not} equivalent to $\mathbf{k}$ as per Eq.~(\ref{eq:cogroup}). All the vectors in the same star have conjugate little co-groups, i.~e.~
\begin{equation}
\bar{G}_{\mathbf{k}R}=R\bar{G}_\mathbf{k}R^{-1}.\label{eq:cogroupstar}
\end{equation}
{Note that if $\bar{G}_\mathbf{k}$ is a normal subgroup of the point group $\bar{G}$, then it is possible for multiple vectors in $*\mathbf{k}$ to belong to the same labelled $\mathbf{k}$-manifold.} For example, in the honeycomb lattice of graphene, the vectors $K$ [with reduced coordinates $(\frac{1}{3},\frac{1}{3})$] and $K'$ [with reduced coordinates $(\frac{2}{3},\frac{2}{3})$] lie in the same star, and are related by the sixfold rotational symmetry of the lattice. Furthermore, the little co-groups $\bar{G}_K$ and $\bar{G}_{K'}$ are isomorphic normal subgroups of the point group $\bar{G}$ (all subgroups of index $2$ are normal\cite{Serre}), and so both belong to the same $\mathbf{k}$-manifold with coordinates $\{(u,u,0) | 0 < u < 2/3\}$ (here and in the remainder of this work, we give the reduced coordinates of $\mathbf{k}$-vectors in the notation of Ref.~\onlinecite{cdml}.)

We say that a {manifold of} vectors $\mathbf{k}$ in the reciprocal lattice is of \emph{maximal symmetry}, if its little co-group is not a subgroup of the little co-group of another {manifold of} vectors $\mathbf{k}'$ connected to it. We shall see that the set of maximal $\mathbf{k}$ for each space group plays a special role in determining the connectivity of energy bands. {We refer to a vector contained in a $\mathbf{k}$-manifold of maximal symmetry as a \emph{maximal $\mathbf{k}$-vector}, or analogously as a $\mathbf{k}$-vector of maximal symmetry.} Note that m$\mathbf{k}$-manifolds of maximal symmetry need not be points; for example in space groups with only a single rotation axis (such as $P6mm$, the space group of $AA$ stacked graphene in an external $z$-directed electric field), $\mathbf{k}$-manifolds are lines directed along the rotation axis when time-reversal symmetry is neglected.

The {non-equivalent (per the equivalence Eq.~\ref{eq:connected_ks} under reciprocal lattice translations) manifolds} of $\mathbf{k}$-vectors for every space group have been tabulated (see for instance Ref.~\onlinecite{cdml}). An on-line database of {the $\mathbf{k}$-vectors for every space group, both maximal and non-maximal,} is accessible via Ref.~\onlinecite{progkvecs}. As discussed in Ref.~\onlinecite{grouptheory}, the set of maximal $\mathbf{k}$-vectors is different depending on whether or not time-reversal (TR) symmetry is considered. In some lines of $\mathbf{k}$-vectors in polar space groups, for example, all the points have the same little co-group and -- as a consequence -- the same symmetry properties (all irreps of the little group {depend smoothly on the coordinate} along the line). However, when TR is considered as an extra (antiunitary) symmetry operation, some points in the line (the $\Gamma$ point and points at the boundary of the first Brillouin zone) are TR-invariant; we refer to these as Time Reversal Invariant Momentum (TRIM) points. At the TRIM points then, TR symmetry sometimes forces irreps that in principle correspond to different energy levels without TR to become degenerate. {A similar issue arises in body- and face-centered space groups at points with antiunitary operations combining a rotation or reflection with TR. Aside from these caveats, our definition of maximal $\mathbf{k}$-vector coincides with the colloquial notion of a ``high-symmetry'' $\mathbf{k}$-vector.}

\begin{figure}
\includegraphics[width=0.4\textwidth]{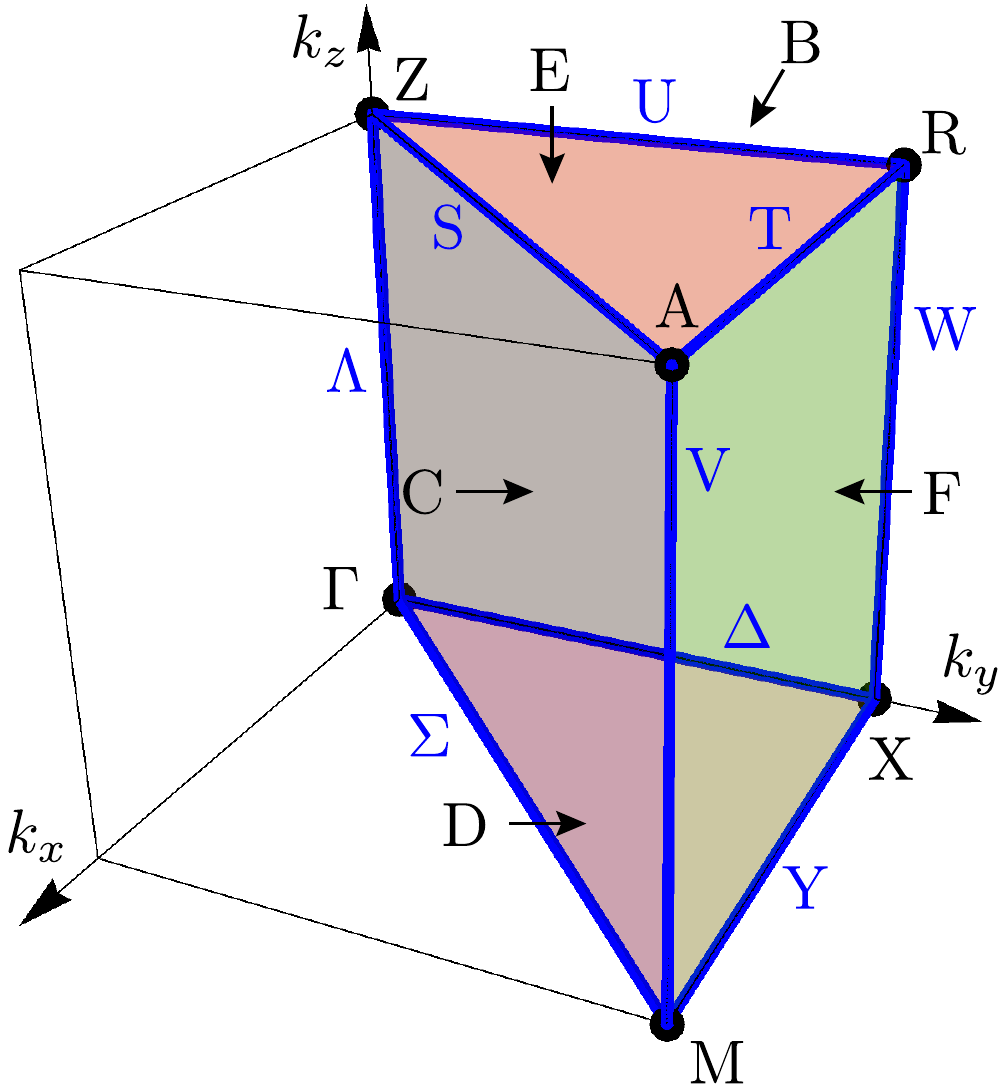}
\caption{Partial view ($0\le k_x,k_y,k_z\le 1/2$ region) of the first Brillouin zone of the space group $P4/ncc$ ($130$). The special $\mathbf{k}$-vectors of Table \ref{table:maximalkvecs}, the points of maximal symmetry $\Gamma$, $Z$, $M$, $A$, $R$, $X$, lines $\Lambda$, $V$, $W$, $\Sigma$, $S$, $\Delta$, $U$, $Y$, $T$ and the planes $D$,  $E$,  $C$, $B$,  $F$ have been indicated.}
\label{fig:brillouin}
\end{figure}

As an example, we give in Table \ref{table:maximalkvecs} the list of $\mathbf{k}$-vectors in the space group $P4/ncc$ ($130$), {sorted into labelled manifolds sharing the same little co-group. This is a tetragonal, non-symmorphic space group, generated by inversion $\{I|000\}$, a fourfold $z$-axis rotation $\{C_{4z}|\half00\}$,  and a two-fold screw rotation $\{C_{2y}|0\half \half\}$ about the $y$-axis.} The first three columns of the table show the label {of the manifold containing each} $\mathbf{k}$-vector, the multiplicity or the number of vectors in its star, and the coordinates of a representative vector of the star in the standard setting, respectively. In the fourth column we give the symbol of the little co-group of each $\mathbf{k}$-{manifold}. In the fifth column, we indicate whether or not each {$\mathbf{k}$-manifold} is maximal. The last column indicates if the TR operator keeps $\mathbf{k}$ invariant. Being a centrosymmetric space group, the set of maximal $\mathbf{k}$-vectors is the same with or without TR: {adding time-reversal is equivalent to adding the composite of inversion and time reversal to the little co-group of every $\mathbf{k}$-vector. Since this does not change the group-subgroup relation of connected $\mathbf{k}$-vectors, it does not change the set of maximal $\mathbf{k}$-vectors as per our definition.} Figure \ref{fig:brillouin} shows the region $0\le k_x,k_y,k_z\le1/2$ of the first Brillouin zone, where the special $\mathbf{k}$-vectors of Table~\ref{table:maximalkvecs} have been indicated.

After having determined all the maximal $\mathbf{k}$-vectors in a given space group (in the following we denote them as $\mathbf{k}^M$), we next compute all the possible connections between each maximal $\mathbf{k}^M$ and all the non-maximal $\mathbf{k}$-vectors. {Each manifold of non-maximal $\mathbf{k}$-vectors is parametrized by} 1 (lines), 2 (planes) or 3 (the general $\mathbf{k}$-vector) free parameters. Note that, to get all the possible connections, we must consider an equation analogous to Eq.~(\ref{eq:connected_ks}) for each vector in $*\mathbf{k}$. Continuing with our example, in Table \ref{table:connections} we show all the connections between the $\mathbf{k}^M$-vectors and the $\mathbf{k}$-vectors of non-maximal symmetry in the space group $P4/ncc$ ($130$). The first column shows the list of maximal vectors $\mathbf{k}^M$, the second gives the non-maximal $\mathbf{k}$-{manifolds} (lines or planes) connected to each $\mathbf{k}^M$, the third column shows the specific values of the continuous parameters for which the points are connected, and the last column indicates the number of vectors $*\mathbf{k}$ connected to each $\mathbf{k}^M$, {equal to the quotient $|*\mathbf{k}|/|*\mathbf{k}^M|$ (where $|\cdot |$ denotes the number of elements in a set)}. For instance, the four vectors of the star $\mathbf{k}=(0,v,0)\in\Delta$ are $*\mathbf{k}=\{(0,v,0),(0,-v,0),(v,0,0),(-v,0,0)\}$, and are connected to $\Gamma:(0,0,0)$ for $v\rightarrow0$. We have suppressed the trivial connections between the $\mathbf{k}^M$-vectors and the general position $GP=\{(u,v,w)\}$.

{Let us define the set of direct paths that join two maximal $\mathbf{k}$-vectors $\mathbf{k}^M_1$ and $\mathbf{k}^M_2$ as the intersection of the sets of non-maximal $\mathbf{k}_i$ connected to $\mathbf{k}^M_1$ and $\mathbf{k}^M_2$. Using the list of possible connections between $\mathbf{k}$-manifolds in a space group, we can construct the set of all direct paths between pairs of maximal $\mathbf{k}$-vectors.} In our example of space group $P4/ncc$ ($130$), we can construct the set of all direct paths by taking intersections of the sets of connections given in Table \ref{table:connections}. Table \ref{table:paths} shows the result of this analysis. The first and fourth columns give all the pairs of maximal $\mathbf{k}$-vectors. The second column shows the possible direct paths that connect the two $\mathbf{k}^M$-vectors. The third column gives the number of vectors in the star of the intermediate $\mathbf{k}$-vectors of non-maximal symmetry connected to both $\mathbf{k}^M$-vectors. As in Table~\ref{table:connections}, the trivial connection through the general position $GP$, common to all pairs of $\mathbf{k}^M$-vectors, has been omitted in the table.

\begin{table}
{\begin{tabular}{cccccc}
$\mathbf{k}$-vec&mult.&coordinates&little&maximal&TR\\
&&&&co-group\\
\hline
$\Gamma$&1&$(0,0,0)$&$4/mmm$($D_{4h}$)&yes&yes\\
$Z$&1&$(0,0,1/2)$&$4/mmm$($D_{4h}$)&yes&yes\\
$M$&1&$(1/2,1/2,0)$&$4/mmm$($D_{4h}$)&yes&yes\\
$A$&1&$(1/2,1/2,1/2)$&$4/mmm$($D_{4h}$)&yes&yes\\
$R$&2&$(0,1/2,1/2)$&$mmm$($D_{2h}$)&yes&yes\\
$X$&2&$(0,1/2,0)$&$mmm$($D_{2h}$)&yes&yes\\
$\Lambda$&2&$(0,0,w),0<w<1/2$&$4mm$($C_{4v}$)&no&no\\
$V$&2&$(1/2,1/2,w),0<w<1/2$&$4mm$($C_{4v}$)&no&no\\
$W$&4&$(0,1/2,w),0<w<1/2$&$mm2$($C_{2v}$)&no&no\\
$\Sigma$&4&$(u,u,0),0<u<1/2$&$mm2$($C_{2v}$)&no&no\\
$S$&4&$(u,u,1/2),0<u<1/2$&$mm2$($C_{2v}$)&no&no\\
$\Delta$&4&$(0,v,0),0<v<1/2$&$mm2$($C_{2v}$)&no&no\\
$U$&4&$(0,v,1/2),0<v<1/2$&$mm2$($C_{2v}$)&no&no\\
$Y$&4&$(u,1/2,0),0<u<1/2$&$mm2$($C_{2v}$)&no&no\\
$T$&4&$(u,1/2,1/2),0<u<1/2$&$mm2$($C_{2v}$)&no&no\\
$D$&8&$(u,v,0),0<u<v<1/2$&$m$($C_s$)&no&no\\
$E$&8&$(u,v,1/2),0<u<v<1/2$&$m$($C_s$)&no&no\\
$C$&8&$(u,u,w),0<u<w<1/2$&$m$($C_s$)&no&no\\
$B$&8&$(0,v,w),0<v<w<1/2$&$m$($C_s$)&no&no\\
$F$&8&$(u,1/2,w),0<u<w<1/2$&$m$($C_s$)&no&no\\
$GP$&16&$(u,v,w),0<u<v<w<1/2$&1(1)&no&no
\end{tabular}}
\caption{$\mathbf{k}$-vectors of the $P4/ncc$ ($130$) (ordinary, or double) space group with TR symmetry. The first column gives the symbol of the {$\mathbf{k}$-manifold}. The second column gives the number of vectors in the star of a vector in the {$\mathbf{k}$-manifold}. The third column shows the coordinates of one representative $\mathbf{k}$-vector {in the manifold}. The fourth column gives the little co-group in the Hermann-Mauguin(Sch\"onflies) notation. In the fifth column we indicate if the $\mathbf{k}$-{manifold} is maximal. Finally the last column indicates if the $\mathbf{k}$-vector{s in the manifold are TRIMs.}}
\label{table:maximalkvecs}
\end{table}

\begin{table}
\begin{tabular}{llll}
maximal&connected&specific&connections\\
$\mathbf{k}$-vec&$\mathbf{k}$-vecs&coordinates&with the star\\
\hline
$\Gamma:(0,0,0)$&$\Lambda:(0,0,w)$&$w=0$&2\\
&$\Delta:(0,v,0)$&$v=0$&4\\
&$\Sigma:(u,u,0)$&$u=0$&4\\
&$B:(0,v,w)$&$v=w=0$&8\\
&$C:(u,u,w)$&$u=w=0$&8\\
&$D:(u,v,0)$&$u=v=0$&8\\
\hline
$Z:(0,0,1/2)$&$\Lambda:(0,0,w)$&$w=1/2$&2\\
&$S:(u,u,1/2)$&$u=0$&4\\
&$U:(0,v,1/2)$&$v=0$&4\\
&$B:(0,v,w)$&$v=0,w=1/2$&8\\
&$C:(u,u,w)$&$u=0,w=1/2$&8\\
&$E:(u,v,1/2)$&$u=v=0$&8\\
\hline
$M:(1/2,1/2,0)$&V:$(1/2,1/2,w)$&$w=0$&2\\
&$\Sigma:(u,u,0)$&$u=1/2$&4\\
&$Y:(u,1/2,0)$&$u=1/2$&4\\
&$C:(u,u,w)$&$u=1/2,w=0$&8\\
&$D:(u,v,0)$&$u=v=1/2$&8\\
&$F:(u,1/2,w)$&$u=1/2,w=0$&8\\
\hline
$A:(1/2,1/2,1/2)$&V:$(1/2,1/2,w)$&$w=1/2$&2\\
&$T:(u,1/2,1/2)$&$u=1/2$&4\\
&$S:(u,u,1/2)$&$u=1/2$&4\\
&$C:(u,u,w)$&$u=w=1/2$&8\\
&$E:(u,v,1/2)$&$u=v=1/2$&8\\
&$F:(u,1/2,w)$&$u=w=1/2$&8\\
\hline
$R:(0,1/2,1/2)$&$T:(u,1/2,1/2)$&$u=0$&2\\
&$U:(0,v,1/2)$&$v=1/2$&2\\
&$W:(0,1/2,w)$&$w=1/2$&2\\
&$B:(0,v,w)$&$v=w=1/2$&4\\
&$F:(u,1/2,w)$&$u=0,w=1/2$&4\\
&$E:(u,v,1/2)$&$u=0,v=1/2$&8\\
\hline
$X:(0,1/2,0)$&$\Delta:(0,v,0)$&$v=1/2$&2\\
&$W:(0,1/2,w)$&$w=0$&2\\
&$Y:(u,1/2,0)$&$u=0$&2\\
&$B:(0,v,w)$&$v=1/2,w=0$&4\\
&$F:(u,1/2,w)$&$u=w=0$&4\\
&$D:(u,v,0)$&$u=0,v=1/2$&8
\end{tabular}
\caption{List of $\mathbf{k}$-vectors (second column) connected to each maximal $\mathbf{k}$-vector (first column) in the $P4/ncc$ ($130$) (ordinary, or double) space group. The third column gives the specific values taken by the continuous parameters in the coordinate triplets of the non-maximal $\mathbf{k}$-vecs in column two. The last column indicates how many vectors in the star of the non-maximal $\mathbf{k}$-vectors of the second column are connected to the maximal $\mathbf{k}$-vector. For example, the $\Gamma$ point is connected to the four vectors of the star of $\mathbf{k}=(0,v,0)\in\Delta$: $*\mathbf{k}=\{(0,v,0)$, $(0,-v,0)$, $(v,0,0),(-v,0,0)|v\in[0,\half]\},$ as $v\rightarrow0$.}
\label{table:connections}
\end{table}

\begin{table}
\begin{tabular}{cccc}
1$^{\mathrm{st}}$ maximal&intermediate&connections&2$^{\mathrm{nd}}$ maximal\\
$\mathbf{k}$-vec&path&with the star&$\mathbf{k}$-vec\\
\hline
$\Gamma:(0,0,0)$&$\boldsymbol{\Lambda}\mathbf{:(0,0,w)}$&2&$Z:(0,0,1/2)$\\
&$B:(0,v,w)$&8\\
&$C:(u,u,w)$&8\\
\hline
$\Gamma:(0,0,0)$&$\boldsymbol{\Sigma}\mathbf{:(u,u,0)}$&4&$M:(1/2,1/2,0)$\\
&$C:(u,u,w)$&8\\
&$D:(u,v,0)$&8\\
\hline
$\Gamma:(0,0,0)$&$C:(u,u,w)$&8&$A:(1/2,1/2,1/2)$\\
\hline
$\Gamma:(0,0,0)$&$B:(0,v,w)$&4&$R:(0,1/2,1/2)$\\
\hline
$\Gamma:(0,0,0)$&$\boldsymbol{\Delta}\mathbf{:(0,v,0)}$&2&$X:(0,1/2,0)$\\
&$B:(0,v,w)$&4\\
&$D:(u,v,0)$&4\\
\hline
$Z:(0,0,1/2)$&$C:(u,u,w)$&8&$M:(1/2,1/2,0)$\\
\hline
$Z:(0,0,1/2)$&$\boldsymbol{S}\mathbf{:(u,u,1/2)}$&4&$A:(1/2,1/2,1/2)$\\
&$C:(u,u,w)$&8\\
&$E:(u,v,1/2)$&8\\
\hline
$Z:(0,0,1/2)$&$\boldsymbol{U}\mathbf{:(0,v,1/2)}$&2&$R:(0,1/2,1/2)$\\
&$B:(0,v,w)$&8\\
&$E:(u,v,1/2)$&8\\
\hline
$Z:(0,0,1/2)$&$B:(0,v,w)$&4&$X:(0,1/2,0)$\\
\hline
$M:(1/2,1/2,0)$&$\boldsymbol{V}\mathbf{:(1/2,1/2,w)}$&2&$A:(1/2,1/2,1/2)$\\
&$C:(u,u,w)$&8\\
&$F:(u,1/2,w)$&8\\
\hline
$M:(1/2,1/2,0)$&$F:(u,1/2,w)$&4&$R:(0,1/2,1/2)$\\
\hline
$M:(1/2,1/2,0)$&$\boldsymbol{Y}\mathbf{:(u,1/2,0)}$&2&$X:(0,1/2,0)$\\
&$D:(u,v,0)$&8\\
&$F:(u,1/2,w)$&4\\
\hline
$A:(1/2,1/2,1/2)$&$\boldsymbol{T}\mathbf{:(u,1/2,1/2)}$&2&$R:(0,1/2,1/2)$\\
&$E:(u,v,1/2)$&8\\
&$F:(u,1/2,w)$&4\\
\hline
$A:(1/2,1/2,1/2)$&$F:(u,1/2,w)$&4&$X:(0,1/2,0)$\\
\hline
$R:(0,1/2,1/2)$&$\boldsymbol{W}\mathbf{:(0,1/2,w)}$&2&$X:(0,1/2,0)$\\
&$B:(0,v,w)$&4\\
&$F:(u,1/2,w)$&4
\end{tabular}
\caption{List of all connections between every pair of maximal $\mathbf{k}$-vectors through intermediate lines or planes in the $P4/ncc$ ($130$) (ordinary, or double) space group. {The first and final column give pairs of maximal $\mathbf{k}$-vectors. The second column gives the label of each $
\mathbf{k}$-manifold connecting the pairs. Finally, the third column gives the number of symmetry-related connections lying in each $\mathbf{k}$-manifold, i.e.~the number of connecting vectors in each star.} The set of connections through the bolded paths defines the whole graph, and determines the set of connections along the non-bolded paths.}
\label{table:paths}
\end{table}

\subsection{Set of independent paths}\label{subsec:independentpaths}
The end goal of enumerating all paths through the Brillouin zone is the determination of all the possible connectivity graphs\cite{GraphTheoryPaper}, with a special focus on the graphs for {the building blocks of band theory, the} elementary band representations\cite{Zak1982,Bacry1988,Michel2001,NaturePaper}. The elementary band representations are representations of infinite dimension that can be expressed -- like any representation of the space group -- as a direct sum of {space group irreps, themselves induced} from irreps of the little group of each $\mathbf{k}$-vector in reciprocal space. 
{Recall that the little group $G_\mathbf{k}$ is the subgroup of the space group $G$ that leaves $\mathbf{k}$ invariant, with the understanding that translations act trivially on $\mathbf{k}$. 
The little co-group $\bar{G}_k$ is then the point group of $G_\mathbf{k}$, and linear representations of the little group that we use here can equally well be viewed as projective representations of the little co-group\cite{bradley})}.  
The multiplicities of each irrep of the little group of every maximal vector $\mathbf{k}^M$ have been calculated for all the elementary band representations\cite{grouptheory}. The multiplicities of the irreps of the little group of $\mathbf{k}$-vecs of non-maximal symmetry can be determined from these via the compatibility relations. The procedure can be briefly described as follows. When two lines of $\mathbf{k}$-vectors intersect at a point (or two planes at a line), this intersection point (line) generically has {higher} symmetry than the points that lie on only one line (plane). The little co-group of the intersection point $\mathbf{k}_s$ is thus generically a supergroup of the little co-group of the line, $\bar{G}_{\mathbf{k}}\subset \bar{G}_{\mathbf{k}_s}$ and the little groups also satisfy, $G_{\mathbf{k}}\subset G_{\mathbf{k}_s}$. The matrices of an irrep $\rho$ of the little group $G_{\mathbf{k}_s}$ associated to the symmetry elements that belong to $G_{\mathbf{k}}$ form a representation of the little group of $\mathbf{k}$, known as the restricted (subduced) representation $\rho\downarrow G_{\mathbf{k}}$. In general, this subduced representation is reducible. The compatibility relations give the decomposition of the irreps of $G_{\mathbf{k}_s}$ into irreps of $G_{\mathbf{k}}$ upon subduction. 

{As an example, we can examine compatibility of little group representations between the little groups $G_\Gamma$ and $G_\Delta$ in our example of space group P$4/ncc$. Because it is located at the origin of the Brillouin Zone, representations of the little group $G_\Gamma$ are insensitive to fractional lattice translations, and so are determined by representations of the little co-group $\bar{G}_\Gamma\approx 4/mmm$. Using data on the Bilbao Crystallographic Server\cite{progrep}, we can focus on the $2D$ representation $\Gamma_5^+$ of $G_\Gamma$, characterized by the representation matrices 
\begin{equation}
D_\mathbf{0}^{\Gamma^+_5}(\{I|000\})=\sigma_0,\;\; D_\mathbf{0}^{\Gamma^+_5}(\{C_{4z}|\half00\})=i\sigma_y,\;\; D_\mathbf{0}^{\Gamma^+_5}(\{C_{2y}|0\half\half\})=\sigma_x,
\end{equation}
where $D_\mathbf{0}^\rho(g)$ signifies the representation matrix of element $g$ in the representation $\rho$ at $\mathbf{k}=0$, and $\sigma_0,\sigma_x,\sigma_y,\sigma_z$ are the Pauli matrices augmented by the two-by-two identity matrix $\sigma_0$. If we now subduce this representation onto the little group $G_\Delta\subset G_\Gamma$, we see that $G_\Delta$ contains only $\{C_{2y}|0\half\half\}$ and $\{m_z|\half\half0\}=\{C_{4z}|\half00\}^2$\{I|000\}. We find then that at $\mathbf{k}=0$ (where $\Gamma$ and $\Delta$ are connected) that the subduced representation $\eta=\Gamma^+_5\downarrow G_\Delta$ is determined by
\begin{equation}
D_\mathbf{0}^{\eta}(\{C_{2y}|0\half\half\})=\sigma_x,\;\; D_\mathbf{0}^{\eta}(\{m_z|\half\half0\})=-\sigma_0.
\end{equation}
This is a reducible representation; it decomposes as a direct sum of two representations $\Delta_2\oplus\Delta_3$ with representation matrices
\begin{align}
D_\mathbf{0}^{\Delta_2}(\{C_{2y}|0\half\half\})&=1,\;\; D_\mathbf{0}^{\Delta_2}(\{m_z|\half\half0\})=1 \\
D_\mathbf{0}^{\Delta_3}(\{C_{2y}|0\half\half\})&=-1,\;\; D_\mathbf{0}^{\Delta_3}(\{m_z|\half\half0\})=-1.
\end{align}
Thus we deduce
\begin{equation}
\Gamma^+_5\downarrow G_\Delta\approx \Delta_2\oplus\Delta_3
\end{equation}
}

{Returning to our general considerations, let us consider} two maximal vectors $\mathbf{k}^{M_1}$ and $\mathbf{k}^{M_2}$ connected through the line or plane $\mathbf{k}$, and a given elementary band representation that subduces into the sets of irreps $\{\rho^{M_1}_i\}$, $\{\rho^{M_2}_i\}$ and $\{\rho_i\}$, of the little groups of $\mathbf{k}^{M_1}$, $\mathbf{k}^{M_2}$ and $\mathbf{k}$, respectively. Then the compatibility relations give, on the one hand, the relations between the sets of irreps $\{\rho^{M_1}_i\}\to\{\rho_i\}$ and, on the other hand, between the sets of irreps $\{\rho^{M_2}_i\}\to\{\rho_i\}$. As the electronic bands are continuous functions in reciprocal space, the set of irreps $\{\rho^{M_1}_i\}$ must be connected to the set of irreps at $\{\rho^{M_2}_i\}$ through the $\{\rho_i\}$. The compatibility relations at both endpoints of the connection restrict the different possible ways to connect the irreps at $\mathbf{k}^{M_1}$ and $\mathbf{k}^{M_2}$. Every set of connections between every pair of maximal $\mathbf{k}$-vectors that fulfill the compatibility relations defines a valid band structure, and hence a valid connectivity graph\cite{GraphTheoryPaper}. However, it is not necessary to consider the compatibility relations along \emph{all} the possible intermediate paths that connect a pair of maximal $\mathbf{k}$-vectors. In most cases, there are redundancies in the restrictions imposed by the compatibility relations along different paths. In order to arrive at a computationally tractable problem, we must minimize the set of paths considered in our calculations. In the following, we explain the different sources of redundant connections, and the algorithms used to remove them. We will continue to make use of the practical example of the space group $P4/ncc$ ($130$) and Table \ref{table:paths} to support our explanations.
\begin{enumerate}
\item \textbf{Paths that are a subspace of other paths}

If two maximal $\mathbf{k}$-vectors $\mathbf{k}^M_1$ and $\mathbf{k}^M_2$ are connected both through a plane $\mathbf{k}_p$ and through a line $\mathbf{k}_l$ contained in the plane, the set of compatibility relations between $\mathbf{k}^M_1,\mathbf{k}^M_2$ and $\mathbf{k}_p$ are redundant and can be omitted in the analysis.

To see this, note that as the line $\mathbf{k}_l$ is contained in the plane, $\mathbf{k}_p$, there exist compatibility relations between the irreps of the little group $\mathbf{G}_{\mathbf{k}_l}$ of $\mathbf{k}_l$ and the little group $\mathbf{G}_{\mathbf{k}_p}$ of $\mathbf{k}_p$, $\{\rho^{l}_i\}\to\{\rho^{p}_i\}$. Then the set of compatibility relations between each of $\mathbf{k}^M_1$, $\mathbf{k}^M_2$ and the plane $\mathbf{k}_p$ are completely determined from the set of compatibility relations between $\mathbf{k}^M_1,\mathbf{k}^M_2$ and $\mathbf{k}_l$ and between this line and $\mathbf{k}_p$; formally we have 
\begin{equation}
\rho^M_i\downarrow G_{\mathbf{k}_p}\approx (\rho^M_i\downarrow G_{\mathbf{k}_l})\downarrow G_{\mathbf{k}_p}
\end{equation}
for each representation $\rho^M_i$ of the little groups $G_{\mathbf{k}^M_1}$ and $G_{\mathbf{k}^M_2}$. Thus, compatibility along the plane $\mathbf{k}_p$ places no additional restrictions on connectivity beyond those obtained from the line $\mathbf{k}_l$.

In the example of Table \ref{table:paths}, for instance, we see that along the $\Gamma\leftrightarrow Z$ connection, the line $\Lambda$ is contained in both the $B$ and $C$ planes. Thus, we can neglect the planes $B$ and $C$ in our analysis of connectivity, without loss of generality. More generally, all the connections through planes can be discarded, when there is also a line connecting the same pair of $\mathbf{k}^M$-vectors. Note that the $C$ plane connecting the points $\Gamma$ and $A$ is the only direct connection between these two $\mathbf{k}$-vectors. This means that the symmetry constraints along $\Gamma-C-A$ are independent of the constraints arising from connections through multiple lines in Table~\ref{sec:paths}, and so this connection cannot be discarded.

\item \textbf{Paths related by a symmetry operation}

Let $R$ be a rotational element that belongs to the little co-groups of $\mathbf{k}^{M_1}$ and $\mathbf{k}^{M_2}$, but that does not belong to the little co-group of an intermediate line $\mathbf{k}$ that is connected to both $\mathbf{k}^{M_1}$ and $\mathbf{k}^{M_2}$. These two $\mathbf{k}^{M_i}$ are also connected through the line $\mathbf{k}R$. The sets of compatibility relations between each 
$\mathbf{k}^{M_i}$ and $\mathbf{k}R$ {differ from the compatibility relations between $\mathbf{k}^{M_i}$ and $\mathbf{k}$ by conjugation by $R$\footnote{Note then that for $\bar{G}_\mathbf{k}$ a normal subgroup of $\bar{G}$, the compatibility relations are identical.}}, but a set of connections between the irreps of the little groups of $\mathbf{k}^{M_1}$ and $\mathbf{k}^{M_2}$ through $\mathbf{k}$ uniquely determines the connections through all of the $\mathbf{k}R$ by symmetry. Therefore, a single line or plane of the star of $\mathbf{k}$ gives all the independent restrictions on the connectivity graphs. In the example of Table \ref{table:paths}, the $\mathbf{k}^M$-vectors $Z$ and $A$ are connected through four lines of the star of $S:(u,u,1/2),(-u,u,1/2),(u,-u,1/2)$ and $(-u,-u,1/2)$ but it is only necessary to consider one of them. {Note that since different representatives of the star are chosen independently for each connection, this set of paths may not lie in a single representation domain (i.e.~ a submanifold $M$ of the Brillouin zone such that $M$ contains one $\mathbf{k}$-vector from every star, and $M\bar{G}\cap M =\emptyset$\cite{bradley}). (Note that a similar construction was used by Kane and Mele to define the $\mathbb{Z}_2$ invariant for topological insulators by considering one half of the Brillouin zone which does not map to itself under time-reversal\cite{kanemele05}).}

\item \textbf{Paths that are a combination of other paths}

Let $\mathbf{k}^{M_1}$, $\mathbf{k}^{M_2}$  and $\mathbf{k}^{M_3}$ be three maximal $\mathbf{k}$-vectors. Consider a line $\mathbf{k}_{l_{12}}$ connecting $\mathbf{k}^{M_1}$ and $\mathbf{k}^{M_2}$, a line $\mathbf{k}_{l_{23}}$ connecting $\mathbf{k}^{M_2}$ and $\mathbf{k}^{M_3}$, and a plane $\mathbf{k}_{p_{13}}$ connecting $\mathbf{k}^{M_1}$ and $\mathbf{k}^{M_3}$. If the plane $\mathbf{k}_{p_{13}}$ contains both the lines $\mathbf{k}_{l_{12}}$ and $\mathbf{k}_{l_{23}}$, then set of compatibility relations between the plane and the two maximal $\mathbf{k}^{M_1}$ and $\mathbf{k}^{M_3}$ are not independent from the two sets of compatibility relations obtained from the lines $\mathbf{k}_{l_{12}}$ and $\mathbf{k}_{l_{23}}$. Therefore, the path that includes the plane can be neglected in the analysis of the connectivity graphs. 

For example, in Table \ref{table:paths} the path $Z\leftrightarrow C\leftrightarrow$ M can be omitted because it is possible to define the path
$Z\leftrightarrow S\leftrightarrow A\leftrightarrow V\leftrightarrow M$, 
and the lines $S$ and $V$ are contained in $C$ (see Fig.~\ref{fig:brillouin}). The possible connections between $Z$, $A$ and $M$ along the lines $S$ and $V$ determine the possible connections between $Z$ and $M$ through $C$.
\end{enumerate}

{While the above rules {\bf 1}, {\bf 2}, and {\bf 3} are perhaps obvious from a physical picture of energy bands, we will need to impose them explicitly at the level of our graph algorithms.} We have applied these rules to calculate the independent paths through the Brillouin zone of each space group. In Table~\ref{table:paths}, the nine bolded paths constitute the full set of independent paths to be considered in the analysis of connectivity graphs for the space group $P4/ncc$ ($130$), both with and without TR.

\subsection{Connectivity of non-symmorphic groups.}\label{subsec:nonsymmorphic}

To complete our enumeration of paths through the Brillouin zone {(or more precisely, from Rule \#{\bf 1} above, the representation domain)}, we must now pay special attention to some subtleties that arise for non-symmorphic space groups. The connectivity of energy bands -- and particularly of elementary band representations -- in non-symmorphic space groups was first studied by Michel and Zak \cite{Michel1999,Michel2001}, who pointed out the essential role of the ``\emph{monodromy} of little group representations''\cite{Herring1942} in determining the band connectivity. We briefly review their analysis here, and then adapt it to our algorithm for determining the necessary paths and compatibility through momentum space.\

{\subsubsection{Monodromy of Little Group Representations}\label{sec:monodromy}}

Michel and Zak limited their analysis to spinless (i.e.~single-valued) band representations, for the most part ignoring TR symmetry as well. First, they analyzed the 9 non-symmorphic space groups {(Bieberbach groups\cite{grouptheory})} generated by the lattice translations and a single screw axis or a glide plane: P$c$, C$c$, P$2_1$, P$3_1$, P$3_2$, P$4_1$,  P$4_3$, P$6_1$ and P$6_5$. The only Wyckoff position in these groups is the general one (and so it is, by fiat, a maximal Wyckoff position\cite{NaturePaper}) and the unique irreducible representation of the site-symmetry group (the group which leaves a representative point in this Wyckoff position invariant, which is isomorphic to point group 1, the trivial group in this case) induces a single elementary representation. {If we allow for spin, we find that there is one double-valued (spinor) band representation as well.} In general, if $\{R|\mathbf{t}\}$ represents a screw axis or a glide plane, and ${n\in\{2,3,4,6\}}$ is the order of $R$, then $\{R|\mathbf{t}\}^n=\{E|p\mathbf{T_R}\}$, where $\mathbf{T_R}$ is a lattice translation, and $p$ is an integer that satisfies $0<p<n$. {We will characterize a non-symmorphic space group by first finding the set of elements with the largest $n$, and from among those choosing the smallest $p$.} If we consider the space groups with screw rotations from the previous list, we {see that for the largest order $n$ of operations in the point group, it is possible to choose one such} that $p=1$. Thus, for these space groups we can take as basis vectors of the lattice of translations $\mathbf{e}_3=\mathbf{T}_R$ and two vectors orthogonal to the screw axis, $\mathbf{e}_1$ and $\mathbf{e}_2$. The corresponding basis vectors of the reciprocal lattice are those vectors $\mathbf{e}_i^*$ that satisfy $\mathbf{e}_i\cdot\mathbf{e}_j^*=2\pi\delta_{ij}$. {For example, the space group P$6_1$ contains the sixfold screw $\{C_{6z}|00\frac{1}{6}\}$, which has $(n,p)=(6,1)$. Similarly, if we consider the space group P$6_5$, this contains the operation $\{C_{6z}^{-1}|00\frac{1}{6}\}$, which also satisfies $(n,p)=(6,1)$}. 

For $\mathbf{k}$-vectors of the form $\mathbf{k}=k_3\mathbf{e}_3^*$, the little group $G_{\mathbf{k}}$ is the whole space group $G$, and its $n$ irreps (before considering TR symmetry at $k=0,1/2$) are one-dimensional. The $n$ {explicitly ($\mathbf{k}$-dependent, from the fractional lattice translation)} one-dimensional matrices {$D_{\mathbf{k}}^{\rho_j}(\{R|\mathbf{t}\})$} of the symmetry operation $\{R|\mathbf{t}\}$ in these simple irreps $\rho_j$, $j=0,1,\ldots,n-1$ are
\begin{equation}
D_{\mathbf{k}}^{\rho_j}(\{R|\mathbf{t}\})=e^{i(2\pi j+k_3)/n}.
\end{equation}
When $k_3$ varies in a full period, $k_3\to k_3+1$, we see that
\begin{equation}
D_{\mathbf{k}}^{\rho_j}(\{R|\mathbf{t}\})\to D_{\mathbf{k}}^{\rho_{j+1}}(\{R|\mathbf{t}\})
\end{equation}
Thus, as we move along the $\mathbf{e}_3^*$ direction in reciprocal space, there is a cyclic permutation of the $n$ irreps of $G_{\mathbf{k}}$. This phenomenon is called monodromy\cite{Herring1942}. Since the Bloch wavefunctions, and hence the energy bands, are periodic functions in the BZ, we deduce as a consequence of the cyclic permutation of irreps along $\mathbf{e}_3^*$ that all $n$ irreps must be connected. This monodromy property also holds for the two space groups with a glide plane, P$c$ and C$c$, {as is evident by writing a glide reflection as the composition of inversion and a twofold screw rotation; $n=2$ for these cases. Thus, monodromy ensures that all elementary band representations (without spin or time reversal) are connected for the $9$ nontrivial Bieberbach groups listed above}. 

Slightly different arguments were used in Ref.~\onlinecite{Michel1999} to demonstrate the full connectivity of elementary band representations in the 5 {additional} space groups P$4_2$, I$4_1$, P$6_3$, P$6_2$ and P$6_4$, for which $p>1$, {which are not Bieberbach groups}. Note that monodromy acts as a cyclic permutation of order 2 in the first three of these groups, and of order 3 in the last two groups. Therefore, not all $n$ irreps are forced to be connected by monodromy alone. In the first three groups the irreps are necessarily connected pairwise, and in the last two groups there are sets of three irreps internally connected. {Finally,} Michel and Zak {explicitly computed the band representations in these groups to}  demonstrate that the irreps not connected by monodromy never occur in the same elementary band representations, {and hence deduce the full connectivity of the (spinless) elementary band representations in these 5 groups.}
Finally Michel and Zak analysed explicitly the connectivities of the space groups I$2_12_12_1$ and I$2_13$, and directly proved the connectivity of their elementary band representations\cite{Michel1999}.

These results, and the fact that the 14=9+5 space groups of the previous first two lists contain all types of {screw rotations and glide reflections} operations in all the 230 space groups, led the authors to conclude that \emph{all} the (spinless) elementary band representations in non-symmorphic space groups are connected, i.~e.~that one can travel continuously through all energy bands. However, we have shown in certain cases that this extrapolation is not justified, due to three possibilities that were not previously considered. First, in some space groups, for example {in certain elementary band representations of the space groups} P$4_2/m$, $I4_1/a$ or I$\bar{4}c2$, the multiplicity of the irreps permuted under the monodromy operation is $2$ or $3$, {rather than $4$}. It is thus possible in these cases to separate the irreps into disconnected subsets where irreps permute amongst themselves under monodromy, each irrep occuring only once per subset. 
An additional subtlety is that, for space groups other than $P6_3$ which contain $6_3$ type screws {(i.e.~screws of the form $\{C_{6}|\frac{1}{2}\mathbf{T}_R\}$)}, the special logic (recapitulated above) that Michel and Zak used to prove the connectivity breaks down -- in some of these groups, little group representations which are not connected by monodromy do in fact appear in the same elementary band representation. In fact, we have found that in space groups P$6_3/m$, P$6_3/mcm$ and P$6_3/mmc$, some elementary band representations can be disconnected. Finally, due to the glide planes in the space groups P$c$ and C$c$ the two irreps at $\Gamma$ are connected. But in other space groups with glide planes, there are more than 2 irreps in the little groups of lines contained in the glide planes as, for example, in the space group P$4/ncc$. Then the necessary pairwise connections of irreps does not guarantee that all the irreps are connected. {For example, let us focus on the case of space group P$6_3/m$. In particular, the points $\Gamma=(0,0,0)$ and $A=(0,0,\half)$ are connected by the line $\Delta=(0,0,w),\;w\in[0,\half]$. There are twelve one-dimensional single-valued representations of $G_\Gamma$, permuted pairwise under monodromy $w\rightarrow 1-w$ as 
\begin{equation}
(\Gamma^\pm_1\leftrightarrow\Gamma^\pm_2),(\Gamma^\pm_3\leftrightarrow\Gamma^\pm_4),(\Gamma^\pm_5\leftrightarrow\Gamma^\pm_6).\label{eq:p6/m}
\end{equation}
Additionally, there are three single-valued representations of $G_A$ labelled $A_1,A_2,$ and $A_3$, all two-dimensional. Along $\Delta$, the compatibility relations force $A_1$ to connect to $(\Gamma_1^\pm,\Gamma_2^\pm)$, $A_2$ to connect to $(\Gamma_3^\pm,\Gamma_4^\pm)$, and $A_3$ to connect to $(\Gamma_5^\pm,\Gamma_6^\pm)$; we thus deduce that each of the representations $A_1$, $A_2$, and $A_3$ are invariant under monodromy.

Unlike in the space group P$6_3$ considered by Michel and Zak, the space group P$6_3/m$ has elementary band representations with more than two bands. In particular, the elementary band representation induced from the $A^g$ representation of the stabilizer group of the $6g$ Wyckoff position (isomorphic to the group $\bar{1}$) subduces the six representations $\Gamma_1^+,\Gamma_2^+,\Gamma_3^+,\Gamma_4^+,\Gamma_5^+$, and $\Gamma_6^+$ of $G_\Gamma$, and the three representations $A_1$, $A_2$, and $A_3$ of $G_A$. From Eq.~(\ref{eq:p6/m}), we see that these representations can be grouped into three disconnected sets of bands.
}

{Let us now consider more generally the consequences of monodromy in non-symmorphic groups, as it pertains to band connectivity}. In general, if along the  $\mathbf{k}=k_3\mathbf{e}_3^*$ direction there are maximal $\mathbf{k}$-vectors for specific values of $k_3$ (for example the $\Gamma$ point for $k_3=0$ or a point at the boundary of the first Brillouin zone), equivalent points are separated along the line with periodicity one (in units of $\mathbf{e}_3^*$). The compatibility relations at these equivalent points give different relations between the irreps at the maximal $\mathbf{k}$-vectors and a general point in the line. In general, for a screw axis that satisfies the above relation $\{R|\mathbf{t}\}^n=\{E|p\mathbf{T_R}\}$, 
{with $n$ chosen to be as large as possible and subsequently $p$ chosen to be as small as possible, as above,} 
there are $n/p$ distinct sets of compatibility relations at $n/p$ different but equivalent points of the line $\mathbf{k}_p=(k_3+j)\mathbf{e}_3^*.\;j\in\{0,1,\dots,n/p-1\}$. For glide planes there are, in general, $2$ different sets of compatibility relations. However, it can be proved that, for all glides and screws, the maximum number of \emph{independent} sets of compatibility relations is $2$. {To see this, consider two equivalent vectors $\mathbf{k}$ and $\mathbf{k}'=\mathbf{k}+\mathbf{e}_3^*$ on the boundary of the first Brillouin zone. There are two compatibility relations at $\mathbf{k}$ and $\mathbf{k}'$ along the line $k_3\mathbf{e}_3^*$, related by monodromy. However, because energy bands are periodic in the first Brillouin zone (irrespective of representation label), these two compatibility relations must completely determine the band structure. This is a physical manifestation of the fact that the monodromy groups are cyclic, i.~e. generated by a translation by $\mathbf{e}_3^*$ in reciprocal space.}

\begin{figure}[t]
\includegraphics[height=2.5in]{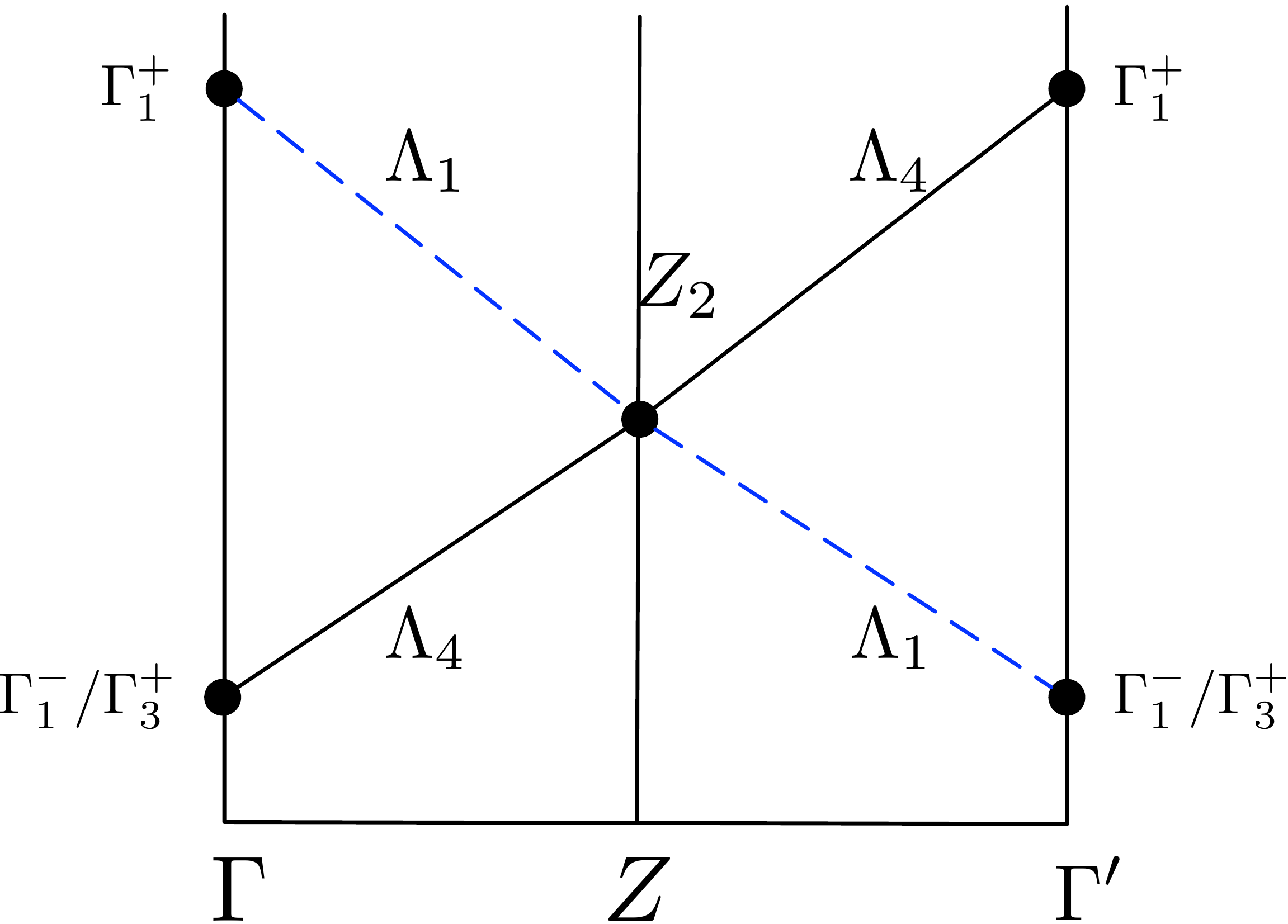}
\caption{{Constraints on energy bands due to glide-reflection symmetry along the $\Gamma-\Lambda-Z$ line in space group P$4/ncc$ ($130$). The points $\Gamma$ and $\Gamma'$ are equivalent, reltated by a translation by the reciprocal lattice vector $\mathbf{K}=\mathbf{e}^*_3$. The solid black band carries the representation $\Lambda_4$, while the dashed blue band carries the representation $\Lambda_1$}. Inversion symmetry pins the monodromy-enforced band crossing at the $Z$ point.}\label{fig:glidebandstructure}
\end{figure}

In our example of the space group P$4/ncc$ ($130$), the lines $\Delta$, $U$, $Y$, $T$, $\Sigma$ and $S$ ({defined in Table~\ref{table:maximalkvecs} and} Fig.~\ref{fig:brillouin}) contain in their little groups $2$-fold screw rotation, and glide reflection: {for example $G_U$ contains the operations $\{C_{2y}|0\half\half\}$ and $\{m_z|\half \half 0\}$. Similarly, the little groups of the lines $\Lambda$, $V$ and $W$ contain the glide reflection $\{m_x|0\half 0 \half\}$}. Therefore, we must consider, in general, $2$ different sets of compatibility relations between {representations of the little groups of} these lines, and {of the little groups of} $2$ distinct maximal $\mathbf{k}$-vectors in the line differing by a reciprocal lattice translation. In some cases the (in-principle) different sets of compatibility relations can be equivalent. This occurs for instance along the $S$ and $\Sigma$ lines in our example: the two sets of compatibility relations at each maximal $\mathbf{k}$-vector ($Z-S$, $A-S$, $\Gamma-\Sigma$ and $M-\Sigma$) are identical. However, in all other lines in this space group, the two sets of compatibility relations are different. For instance, let us consider the path $\Gamma-\Lambda-Z$, {where the relevant little-group operation is the glide reflection $\{m_x|\half 0 \half\}$}. The compatibility relations for irreps between $Z:(0,0,1/2)$ and $\Lambda:(0,0,w)$ are:
\begin{equation}\label{eq:comprelformat}
\begin{array}{lcl}
Z_1&\to&\Lambda_2\oplus \Lambda_3\\
Z_2&\to&\Lambda_1\oplus \Lambda_4\\
Z_3&\to&\Lambda_5\\
Z_4&\to&\Lambda_5
\end{array}
\end{equation}
The same set of compatibility relations are obtained at all the points equivalent to $Z$ in the line $\Lambda$. {A shift of $\mathbf{k}$ by $\mathbf{e}^*_3$ interchanges the representations $\Lambda_2\leftrightarrow\Lambda_3$, $\Lambda_1\leftrightarrow\Lambda_4$, and $\Lambda_5\leftrightarrow\Lambda_5$}. However, the compatibility relations at $\Gamma$:(0,0,0) and at the equivalent point $\Gamma$':(0,0,1) are:
\begin{equation}
\label{eq:comprelisomorphic}
{
\begin{array}{lcc|c}
&&\Gamma:(0,0,0)\rightarrow \Lambda:(0,0,w=0)&\Gamma ':(0,0,1)\rightarrow \Lambda:(0,0,w=1)\\
\hline
\Gamma_1^{+}&\to&\Lambda_1&\Lambda_4\\
\Gamma_1^{-}&\to&\Lambda_4&\Lambda_1\\
\Gamma_2^{+}&\to&\Lambda_2&\Lambda_3\\
\Gamma_2^{-}&\to&\Lambda_3&\Lambda_2\\
\Gamma_3^{+}&\to&\Lambda_4&\Lambda_1\\
\Gamma_3^{-}&\to&\Lambda_1&\Lambda_4\\
\Gamma_4^{+}&\to&\Lambda_3&\Lambda_2\\
\Gamma_4^{-}&\to&\Lambda_2&\Lambda_3\\
\Gamma_5^{+}&\to&\Lambda_5&\Lambda_5\\
\Gamma_5^{-}&\to&\Lambda_5&\Lambda_5
\end{array}
}
\end{equation}
Looking at the lists in Eq.~(\ref{eq:comprelisomorphic}) we see, for example, that $\Gamma_1^{+}$ is necessarily connected to either $\Gamma_1^{-}$ or $\Gamma_3^{+}$ and to Z$_2$ at the $Z$ point, but not all the irreps at the $\Gamma$ point are necessarily connected due to the glide plane parallel to the $\Lambda$ line. {We show this graphically for the band structure along the $\Gamma-\Lambda-Z-\Lambda-\Gamma'\equiv\Gamma$ line in Fig.~\ref{fig:glidebandstructure}.} In the following, we distinguish both sets of irreps at the intermediate path by a superscript. Taking as an example the irrep $\Gamma_1^{+}$ in Eq.~(\ref{eq:comprelisomorphic}), we write the compatibility relations as $\Gamma_1^{+}\to$ $\Lambda_1^1$ and $\Gamma_1^{+}\to$ $\Lambda_4^2$.

{\subsubsection{Monodromy and the Minimal Set of Paths}\label{subsec:nonsymrule}

Now that we understand the role monodromy plays in enforcing connectivity of energy bands, we can incorporate it into our set of rules from Sec.~\ref{subsec:independentpaths} in order to find the smallest set of non-redundant paths through the Brillouin zone for non-symmorphic space groups. We know from the preceding analysis of Sec.~\ref{sec:monodromy} that for lines $L=\{\mathbf{k}_t\}$ whose little group $G_{L}$ contains either a screw rotation or a glide reflection, that band connectivity along $L$ is fully determined by a pair of distinct compatibility relations for the little group of two equivalent points $G_{\mathbf{k}_t}$ and 
$G_{\mathbf{k}_{-t}+\mathbf{K}}$ in $L$. 

However, referring back to Rule {\bf 1} of Sec.~\ref{subsec:independentpaths}, we notice that these pairs of compatibility relations only impose additional constraints on the band structure when generic points on the line $\mathbf{k}_t$ and $\mathbf{k}_{-t}+\mathbf{K}$ are not related by a symmetry operation, i.e.~when they do not lie in the same star. When the two points \emph{do} lie in the same star, then there exists an element $R$ in the point group such that $\mathbf{k}_tR=\mathbf{K}+\mathbf{k}_{-t}$. The band connectivity at $\mathbf{k}_t$ determines the band connectivity at $\mathbf{k}_{-t}+\mathbf{K}$ under the action of $R$; the monodromy of little group representations then is implemented by conjugation of the little group by $R$, as per Eq.~(\ref{eq:cogroupstar}). In this case, any connectivity constraints enforced by monodromy are restricted by symmetry to occur at maximal $\mathbf{k}$-vectors invariant under the action of $R$.

We can see this cleary in our analysis of the $\Gamma-\Lambda-Z-\Lambda-\Gamma'$ line in space group P$4/ncc$ from Fig.~\ref{fig:glidebandstructure} above. In this space group the two connecting lines $(0,0,v)\in\Lambda$ and $(0,0,1-v)\in\Lambda$ ($v\in[0,\half]$) are related by the action of inversion. Conjugation by inversion interchanges the little group representations $\Lambda_1\leftrightarrow\Lambda_4$, consistent with the monodromy due to the glide-reflection in the little group $G_\Lambda$ (see Fig.~\ref{fig:glidebandstructure}). Consequently, the connectivity of bands implied by this monodromy is forced to occur at the high-symmetry point $Z$, which hosts only two-dimensional representations. The full band connectivity along the line $\Gamma-\Lambda-Z-\Lambda-\Gamma'$ is thus determined uniquely by the connectivity along $\Gamma-\Lambda-Z$, as we can see from the figure. A similar monodromy-enforced band crossing with the addition of TR symmetry was exploited in Ref.~\onlinecite{Hourglass}. 

Taking this into account, we have the additional rule for determining the non-redundant paths through the Brillouin zone for non-symmorphic groups:
\begin{itemize}
	\item  If the little group of a line or plane $\{\mathbf{k}_t\}$ contains a screw rotation or a glide reflection, and if $\mathbf{k}_t$ is not in the star of $K+\mathbf{k}_{-t}$, then we must take into account two sets of compatibility relations for representations of $G_{\mathbf{k}_t}$ related by monodromy.
\end{itemize}

In particular, note that in any space group with TR or inversion symmetry, $-\mathbf{k}\in*\mathbf{k}$ for every $\mathbf{k}$-vector. Hence in these space groups we need only consider a single set of compatibility relations for all connections between TRIM points. In the specific example of space group P$4/ncc$ ($130$), we see then that we need only consider a single set of compatibility relations for all connections, since the little group of all maximal $\mathbf{k}$-vectors contains inversion.
}


\subsection{The graph construction algorithm}\label{sec:maiagraph}

Armed with the {sets of paths} through the Brillouin zone for each space group, we would now like to solve the constraints imposed by the group-theoretic compatiblity relations along those paths. These solutions take the form of groups of connected bands, with bands at high symmetry points and lines transforming {under a direct sum of irreps of the space group}. Any set of such bands can be realized as the spectrum of some local Hamiltonian $H$ (including topological bands, although these cannot appear in isolation). However, since the compatibility relations are a purely group-theoretic device with meaning independent of any choice of Hamiltonian, it is useful for us to devise a more refined graph-theoretic picture of band connectivity. Our final goal is the classification of all valid band structures for all elementary band representations\cite{NaturePaper,GraphTheoryPaper} in each space group. To begin, we first review some graph-theoretic terminology.

\subsubsection{Review of Connectivity Graphs}\label{sec:graphreview}

\begin{figure}[h]
\centering
    \includegraphics[scale=0.3,angle=90]{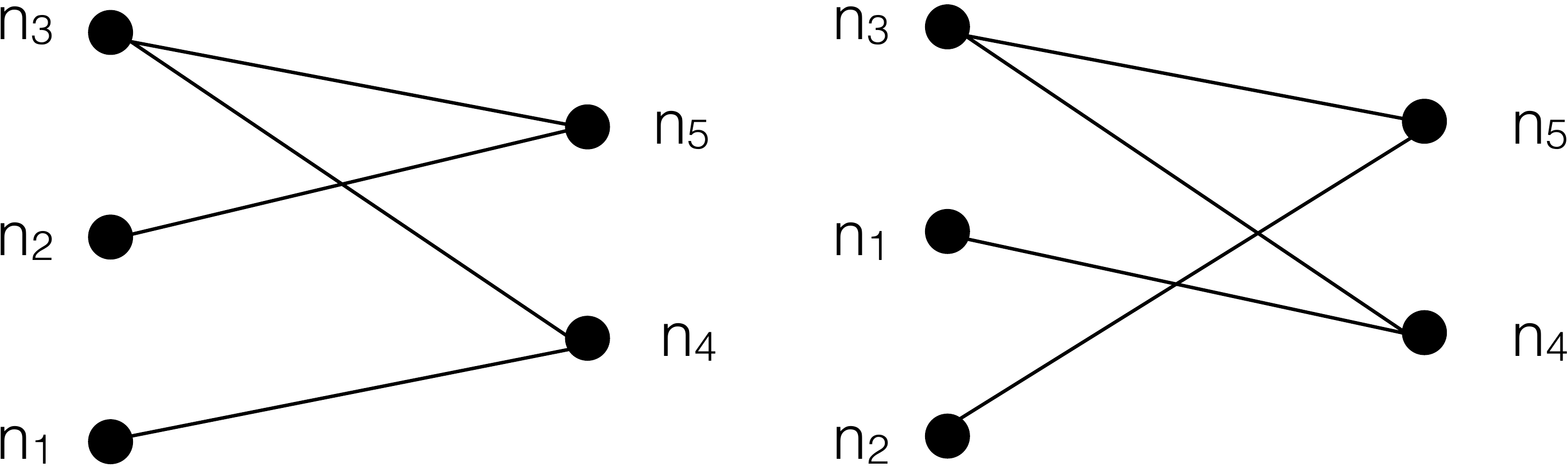}
    \caption{Example of a graph connected in 2 different ways. Solid circles are nodes connected by 4 edges (solid lines). We have 2 partitions, \{n$_1$,n$_2$,n$_3$\} and \{n$_4$,n$_5$\}. We have 2 ways of connecting the different irreps.}
\label{fig:0}        
\end{figure}

We aim to enumerate and classify all the valid band structures consistent with the symmetry constraints in each of the $230$ space groups. In order to ensure that the compatibility relations are satisfied along the non-redundant lines and planes joining all pairs of maximal $\mathbf{k}$-vectors, we will employ the mapping to graph theory {introduced} in Ref.~\onlinecite{NaturePaper}, {and expanded upon in detail in Ref.~}\onlinecite{GraphTheoryPaper}. Here we briefly review the basic concepts of spectral graph theory, and the tools that will be needed for our algorithm. First, recall that a {\bf partition} of a graph is a subset, V$_0$, of nodes such that no two nodes in V$_0$ are connected by an
edge. In our construction, each partition will correspond to a high-symmetry (manifold of) $\mathbf{k}$-vector(s), and irreps of the little group of each $\mathbf{k}$-vector will be represented as nodes (see Fig.~\ref{fig:0}).
Next, the {\bf degree} of a node $p$ in a graph is the number of edges that end on $p$.
Using these ideas, we defined in Refs.~\onlinecite{NaturePaper,GraphTheoryPaper} the connectivity graph for a collection of little group representations $\mathcal{M}$ (i.~e.~bands) forming a (physical) band representation for a space group $G$ as follows:
\begin{defn*}
{Given an elementary band representation with little group representations $\mathcal{M}$, we construct a {\bf connectivity graph} $C_\mathcal{M}$ using the following rules: We associate a node, $n^a_{\mathbf{k}_i}\in C_\mathcal{M}$, in the graph to each representation $\rho_{\mathbf{k}_i}^a\in\mathcal{M}$ of the little group $G_{\mathbf{k}_i}$ of every high-symmetry manifold (point, line, plane, and volume), $\mathbf{k}_i$.}
If an irrep occurs multiple times in $\mathcal{M}$, there is a separate node for each occurence.
{The degree of each node, $n^a_{\mathbf{k}_i}$, is $P_{\mathbf{k}_i}\cdot \mathrm{dim}(\rho^a_{\mathbf{k}_i})$, where $P_{\mathbf{k}_i}$ is the number of high-symmetry manifolds connected to the point $\mathbf{k}_i$}:  $\mathrm{dim}(\rho^a_{\mathbf{k}_i})$ edges lead to each of these other $\mathbf{k}-manifolds$ in the graph, one for each energy band. 
When the manifold corresponding to $\mathbf{k}_i$ is contained within the manifold corresponding to $\mathbf{k}_j$, as in a high-symmetry point that lies on a high-symmetry line, their little groups satisfy $G_{\mathbf{k}_j} \subset G_{\mathbf{k}_i}$. {For each node $n_{\mathbf{k}_i}^a$, we compute
\begin{equation}
\rho_{\mathbf{k}_i}^a\downarrow G_{\mathbf{k}_j}\approx\bigoplus_{b}\rho_{\mathbf{k}_j}^b.
\end{equation}
We then connect each node $n_{\mathbf{k}_j}^b$ to the node $n_{\mathbf{k}_i}^a$ with $\mathrm{dim}(\rho^b_{\mathbf{k}_j})$ edges.
}
\end{defn*}

A solution to the compatibility relations is thus equivalent to a way of connecting nodes to form a connectivity graph consistent with the above definition. Therefore, using the compatibility relations, we will algorithmically construct all valid connectivity graphs for each of the {$5646$} elementary band representations, as well as for the $4757$ independent \emph{physically} elementary band representations, i.~e. those band representations that are elementary in the presence of TR symmetry\cite{NaturePaper}. To find those connectivity graphs that separate into disconnected components, and hence to identify topologically nontrivial groups of isolated bands, we employ some standard techniques of spectral graph partitioning\cite{GraphThy}. In particular, recall that to every graph with $m$ nodes, we can define an $m\times m$ {\bf adjacency matrix} $A$, where the $(ij)$th entry is the number of edges connecting node $i$ to node $j$. A graph is uniquely determined by its adjacency matrix. Furthermore, we introduce the {\bf degree matrix} $D$ of a graph, a diagonal matrix whose $(ii)$th entry is the degree of the node $i$. From these we define the {\bf Laplacian matrix}
\begin{equation}
L\equiv D-A.
\end{equation}
The spectrum of $L$ has the following useful property: For each connected component of a graph $G$, there is a zero eigenvector of the Laplacian of $G$. Furthermore, the components of the eigenvector with this eigenvalue can be chosen to be $1$ on all nodes in the connected component, and $0$ on all others. The proof of this statement is given in Refs.~\onlinecite{GraphThy,GraphTheoryPaper}, and follows directly from the observation that the sum of entries in any row of the Laplacian matrix is by definition zero, coupled with the observation that if $L_{ij}\neq 0$, then nodes $i$ and $j$ lie in the same connected component. 

Next, we will use the facts above to construct our algorithm, with which we have tabulated all realizable topological phases for each elementary band representation. First, we will describe our method for constructing all connectivity graphs for a band representation, given the data of Secs.~\ref{sec:paths}--\ref{subsec:nonsymmorphic}. Then, in Sec.~\ref{sec:maialaplacian}, we will describe some details of the implementation of the spectral graph partitioning method outlined above.

\subsubsection{Direct construction of connectivity graphs}

We now design an algorithm to construct and diagonalize the Laplacian matrix for the connectivity graphs of an elementary band representation, separating the task into two steps. We first construct all possible adjacency matrices, and then we subtract the degree matrix from them to obtain the Laplacian. Note that by the definition of the connectivity graphs, the adjacency matrix has a natural block structure: blocks are indexed by a pair $(\mathbf{k}_1,\mathbf{k}_2)$ of $\mathbf{k}$-vector partitions, and a block contains a nonzero submatrix if and only if those two $\mathbf{k}$-vectors are compatibile (c.~f.~Sec.~\ref{sec:paths}). We exploit this block structure to first build each nonvanishing submatrix separately. We then combine the submatrices together in all distinct ways to form the adjacency matrices for all valid connectivity graphs.

Let us begin with an elementary band representation $\rho\uparrow G$ in a space group $G$. We start by noting that, as per the results of Sec.~\ref{subsec:independentpaths}, we need only concern ourselves with the minimal set of independent paths in the BZ of $G$ to fully constrain the connectivity graphs. As such, our adjacency matrices will have $n_\mathbf{k}\times n_\mathbf{k}$ blocks, where $n_\mathbf{k}$ is the number of $\mathbf{k}$-vectors appearing in the nonredundant paths through the BZ, with appropriate care taken to distinguish independent paths in non-symmorphic groups as per Sec.~\ref{subsec:nonsymmorphic}. Because the adjacency matrix is symmetric with vanishing diagonal blocks (no edges connect nodes in the same partition), we need only analyze the blocks above the diagonal. To this end, we order the blocks of the adjacency matrix by placing the maximal $\mathbf{k}$-vectors first, followed by the non-maximal ones. In this way, all the submatrices which we analyze have rows labelled by irreps of the little group of maximal $\mathbf{k}$-vectors, and columns labelled by irreps of the little group of non-maximal $\mathbf{k}$-vectors. The precise representations are determined by subduction $\rho\uparrow G\downarrow G_{\mathbf{k}}$ of the band representation onto the little groups of each $\mathbf{k}$-vector, as obtained in Ref.~\onlinecite{grouptheory}.

We next construct a valid submatrix making use of the compatibility tables\cite{NaturePaper,grouptheory} along the paths through the BZ. As discussed above, if $\mathbf{k}_1$ and $\mathbf{k}_2$ are not compatible $\mathbf{k}$-vectors, then the only valid submatrix for a block $(\mathbf{k}_1,\mathbf{k}_2)$ is the zero matrix. Let us consider then the block $(\mathbf{k}^{M_1},\mathbf{k}_t)$, with $\mathbf{k}^{M_1}$ a maximal $\mathbf{k}$-vector, and $\mathbf{k}_t$ a nonmaximal vector compatible with it. Then this block is nonzero, and the entries in the submatrix fulfill the following rules: First, we only allow one nonzero entry per column. This reflects the fact that each representation of the little group $G_{\mathbf{k}^{M_1}}$ subduces onto a unique sum of representations of $G_{\mathbf{k}_t}$. Second, the sum of the entries in each row must be the dimension of the corresponding little-group representation. This corresponds to the fact that subduction does not change the dimension of a representation. Given a single valid submatrix, we generate all other valid submatrices by considering column-wise permutations: given a valid submatrix with two columns labelled by isomorphic little group representations, then the submatrix obtained by permuting these columns will also be a valid submatrix.

As an example let us continue with the space group P4$/ncc$ (130).  As shown in Table~\ref{table:paths}, and the discussion in Sec.~\ref{subsec:nonsymmorphic}, the maximal $\mathbf{k}$-vectors are labelled $A$, $\Gamma$, $M$, $R$, $X$ and $Z$. The independent connections are along the $\Lambda$, $\Sigma$, $\Delta$, $S$, $U$, $V$, $Y$, $T$, and $W$ lines/planes. {Recall from Sec.~\ref{subsec:nonsymrule} that even though this space group is nonsymmorphic, we only need one set of compatibility relations along each connection in order to determine the allowed band connectivities}. 
Let us consider the {elementary} band representation induced from the $\bar{A}^u$ ($\bar{\Gamma}_3$) double-valued representation of the site-symmetry group of the {maximal} $8d$ Wyckoff position. {Because the Wyckoff multiplicity of this site is 8, and because the $\bar{A}^u$ representation is one-dimensional, there are 8 bands in this band representation.} For simplicity, we will consider the case without TR symmetry. From Ref.~\onlinecite{grouptheory}, we know that this band representation subduces to the $\bar{A}_5\oplus\bar{A}_5$ representation at the $A$ point, and the $\bar{V}_6\oplus\bar{V}_6\oplus\bar{V}_7\oplus\bar{V}_7\equiv 2\bar{V}_6\oplus2\bar{V}_7$ ({here and throughout, we will indicate the repeated direct sum of a representation with itself by integer multiplication).} representation along the line $V$. The compatibility relations for the little groups $G_A\rightarrow G_{V}$ are
\begin{equation}
\bar{A}_5\rightarrow \bar{V}_6\oplus\bar{V}_7.
\end{equation}

Using these, one valid submatrix for the $(A,V)$ block of the adjacency matrix is shown in Table~\ref{table:SUB1}.
\begin{table}[h]
\begin{tabular}{c|cccc}
  & $\bar{V_6}$ & $\bar{V_6}$ & $\bar{V_7}$ & $\bar{V_7}$\\
 \hline
  $\bar{A_5}$ & 1 & 0 & 1 & 0 \\ 
  $\bar{A_5}$ & 0 & 1 & 0 & 1
  \end{tabular}
\caption{One possible valid submatrix for the $A$-$V$ connection of P4$/ncc$ 8d $\bar{\Gamma}_{3}$ elementary band representation.}
\label{table:SUB1}
\end{table}
By permuting the identically labelled $\bar{V}_6$ and $\bar{V}_7$ columns, we arrive at the following three additional submatrices shown in Table~\ref{tab:othersubs}
\begin{table}[ht]
\begin{tabular}{c|cccc}
  & $\bar{V_6}$ & $\bar{V_6}$ & $\bar{V_7}$ & $\bar{V_7}$\\
 \hline
  $\bar{A_5}$ & 0 & 1 & 1 & 0 \\ 
  $\bar{A_5}$ & 1 & 0 & 0 & 1
  \end{tabular}\quad
  \begin{tabular}{c|cccc}
  & $\bar{V_6}$ & $\bar{V_6}$ & $\bar{V_7}$ & $\bar{V_7}$\\
 \hline
  $\bar{A_5}$ & 1 & 0 & 0 & 1 \\ 
  $\bar{A_5}$ & 0 & 1 & 1 & 0
  \end{tabular}\quad
  \begin{tabular}{c|cccc}
  & $\bar{V_6}$ & $\bar{V_6}$ & $\bar{V_7}$ & $\bar{V_7}$\\
 \hline
  $\bar{A_5}$ & 0 & 1 & 0 & 1 \\ 
  $\bar{A_5}$ & 1 & 0 & 1 & 0
  \end{tabular}
  \caption{The remaining valid submatrices for the  $A$-$V$ connection of P4$/ncc$ 8d $\bar{\Gamma}_{3}$ elementary band representation.}\label{tab:othersubs}
\end{table}

After constructing all nonzero submatrices for all compatible connections, we next build the full adjacency matrix row by row. In doing so, we would like to avoid overcounting configurations that differ only by a relabelling of representations along non-maximal $\mathbf{k}$-vectors, {, such as exchanging the two identical copies of $V_6$ along the line $V$}. For example, consider a high-symmetry line connecting two maximal $\mathbf{k}$-vectors. Connections along this line are specified by two separate blocks in the adjacency matrix, each with its own set of valid submatrices. To ensure we consider only distinct connections along this line, we fix a valid submatrix for one endpoint, thus fixing the connections at one endpoint of the non-maximal line.
Incorporating this constraint into our algorithm, we first sort our maximal $\mathbf{k}$-vectors by number of little group representations, from highest to lowest. Going row-by-row through the adjacency matrix, the first time a non-maximal $\mathbf{k}$-vector appears in a non-vanishing block, we can arbitrarily choose a fixed, valid submatrix for this block. In all subsequent blocks corresponding to this non-maximal $\mathbf{k}$-vector we must consider all allowed valid submatrices. We summarize these rules in Table~\ref{table:AD1}. 
\begin{table}[h!]
\begin{tabular}{c|ccccccccccccccc}
  & $A$ & $\Gamma$ &  $M$ & $R$ & $X$ & $Z$ & $S$ & $V$ & $T$ & $\Delta$ & $\Sigma$ &  $\Lambda$ & $Y$ & $U$ & $W$ \\
 \hline
$A$ &0 & 0 & 0 & 0 & 0 & 0 & F & F & F & 0 & 0 & 0 & 0 & 0 & 0  \\
$\Gamma$  & * &  0 & 0 & 0 & 0 & 0 & 0 & 0 & 0 & F & F & F & 0 & 0 & 0 \\
$M$   & * & * &  0 & 0 & 0 & 0 & 0 & P & P & 0 & P & 0 & F &  0 & 0 \\
$R$  & * & * & * &  0 & 0 & 0 & 0 & 0 & 0 & 0 & 0 & 0 & 0 & F &  F  \\
$X$  & *&  * & * & * &  0 & 0 & 0 & 0 & 0 & P & 0 & 0 & P & 0 & P \\
$Z$ & * & * & * & * & * & 0 & P & 0 & 0 & 0 & 0 & P & 0 & P & 0  \\
$S$  & * & * & * & * & * & * & 0 & 0 & 0 & 0 & 0 & 0 & 0 & 0 & 0  \\ 
$V$& * & * & * & * & * & * & * & 0 & 0 & 0 & 0 & 0 & 0 & 0 & 0 \\
$T$ & * & * & * & * & * & * & * & * & 0 & 0 & 0 & 0 & 0 & 0 & 0  \\
$\Delta$ & * & * & * & * & * & * & * & * & * & 0 & 0 & 0 & 0 & 0 & 0  \\
$\Sigma$  & * & * & * & * & * & * & * & * & * & * & 0 & 0 & 0 & 0 & 0  \\
$\Lambda^1$&  * & * & * & * & * & * & * & * & * & * & * & 0 & 0 & 0 & 0  \\
$Y$ & * & * & * & * & * & * & * & * & * & * & * & * & 0 & 0 & 0\\
$U$ & * & * & * & * & * & * & * & * & * & * & * & * & * & 0 & 0  \\
$W$ & * & * & * & * & * & * & * & * & * & * & * & * & * & * & 0  \\
\end{tabular}
\caption{Schematic plot of the permuting blocks of the group of adjacency matrices of P4$/ncc$ 8d $\bar{\Gamma}_{3}$ EBR. $F$ stands for fixed block and $P$ for permuting block, {as defined in Tables~\ref{table:SUB1} and \ref{tab:othersubs}}. All adjacency matrices are obtained by choosing a single, fixed submatrix for each block labelled $F$, and \emph{all} valid submatrices for the blocks labelled $P$.}
\label{table:AD1}
\end{table}
{As a specific example, let us consider the $\Gamma$-$\Lambda$-$Z$ line in space group $P4/ncc$. We continue to work with the elementary band representation induced from the $\bar{\Gamma}_3$ representation of the site-symmetry group of the $8d$ Wyckoff position. As indicated in Table~\ref{table:AD1}, we can fix the $\Gamma$-$\Lambda$ submatrix to be any solution to the compatibility relations; we choose the submatrix shown in Table~\ref{table:SUB2}
\begin{table}[ht]
\begin{tabular}{c|cccc}
  & $\bar{\Lambda_6}$ & $\bar{\Lambda_6}$ & $\bar{\Lambda_7}$ & $\bar{\Lambda_7}$\\
 \hline
  $\bar{\Gamma_6}$ & 2 & 0 & 0 & 0 \\ 
  $\bar{\Gamma_6}$ & 0 & 2 & 0 & 0\\
  $\bar{\Gamma_7}$ & 0 & 0 & 2 & 0\\
  $\bar{\Gamma_7}$ & 0 & 0 & 0 & 2
  \end{tabular}
\caption{One possible valid submatrix for the $\Gamma$-$\Lambda$ connection in the elementary band representation of space group P4$/ncc$ induced from the $\bar{\Gamma}_{3}$ irrep of the $8d$ position.}
\label{table:SUB2}
\end{table}
On the other hand, for the $\Lambda$-$Z$ connection, according to Table \ref{table:AD1} we consider all possible submatrices, as shown in Table~\ref{tab:othersubs}
\begin{table}[ht]
\begin{tabular}{c|cccc}
  & $\bar{\Lambda_6}$ & $\bar{\Lambda_6}$ & $\bar{\Lambda_7}$ & $\bar{\Lambda_7}$\\
 \hline
   $\bar{Z_5}$ & 0 & 2 & 0 & 0 \\ 
  $\bar{Z_6}$ & 0 & 0 & 0 & 2\\
  $\bar{Z_7}$ & 2 & 0 & 0 & 0\\
  $\bar{Z_8}$ & 0 & 0 & 2 & 0
  \end{tabular}\quad
  \begin{tabular}{c|cccc}
  & $\bar{\Lambda_6}$ & $\bar{\Lambda_6}$ & $\bar{\Lambda_7}$ & $\bar{\Lambda_7}$\\
 \hline
   $\bar{Z_5}$ & 0 & 2 & 0 & 0 \\ 
  $\bar{Z_6}$ & 0 & 0 & 2 & 0\\
  $\bar{Z_7}$ & 2 & 0 & 0 & 0\\
  $\bar{Z_8}$ & 0 & 0 & 0 & 2
  \end{tabular}\quad
  \begin{tabular}{c|cccc}
  & $\bar{\Lambda_6}$ & $\bar{\Lambda_6}$ & $\bar{\Lambda_7}$ & $\bar{\Lambda_7}$\\
 \hline
   $\bar{Z_5}$ & 2 & 0 & 0 & 0 \\ 
  $\bar{Z_6}$ & 0 & 0 & 0 & 2\\
  $\bar{Z_7}$ & 0 & 2 & 0 & 0\\
  $\bar{Z_8}$ & 0 & 0 & 2 & 0
  \end{tabular}\quad
    \begin{tabular}{c|cccc}
  & $\bar{\Lambda_6}$ & $\bar{\Lambda_6}$ & $\bar{\Lambda_7}$ & $\bar{\Lambda_7}$\\
 \hline
   $\bar{Z_5}$ & 2 & 0 & 0 & 0 \\ 
  $\bar{Z_6}$ & 0 & 0 & 2 & 0\\
  $\bar{Z_7}$ & 0 & 2 & 0 & 0\\
  $\bar{Z_8}$ & 0 & 0 & 0 & 2
  \end{tabular}
  \caption{Valid submatrices for the  $Z$-$\Lambda$ connection in the elementary band representation of space group P4$/ncc$ induced from the $\bar{\Gamma}_{3}$ irrep of the $8d$ position.}\label{tab:othersubs}
\end{table}
}

Even with this procedure to eliminate redundancies, the number of raw adjacency matrices that arise from these permutations, may be computationally intractable. {For instance, the example of Table~\ref{table:AD1} defines on the order of $4^{9}$ adjecency matrices (the exponent is the number of permuting blocks).} We implement some additional {\it filters} to avoid redundancies:
\begin{itemize}
\item {\it Multiple-irrep filter:} Consider a maximal $\mathbf{k}$-vector which in a particular band representation contains only multiple copies of the same representation. Then we can fix a valid submatrix for every block corresponding to this $\mathbf{k}$-vector. This is because permutations amongst these identical irreps cannot change the connectivity of the full graph. We illustrate this for the line $A$-$V^1$-$M$ in space group $P4/ncc$ ($130$) in Fig.~\ref{fig:1}.
\item {\it Fake-Weyl filter:} The inverse map from a connectivity graph to a band structure requires a choice of {embedding (in the formal graph-theoretic sense, c.f.~Ref.~\onlinecite{GraphThy}) of the connectivity graph into the Brillouin zone.} In particular, all nodes in the partition labelled by $\mathbf{k}_i$ map onto the manifold  of points $\mathbf{k}_i$ in the BZ. In this embedding, edges of the connectivity graph may cross, corresponding to crossings of bands in the band structure. Generically, crossings along high-symmetry lines are only protected if the two bands carry different representations of the little group of the line. Accidental crossings of identical representations are not stable to perturbations: they will either gap, or in the case of Weyl nodes (which require broken inversion symmetry), they can be pushed away from high-symmetry lines. Because we are interested in classifying generic, stable band structures, we will discount connectivity graphs corresponding to accidental crossings. Therefore, if we have accidental crossings we only need to consider one combination of the sub-matrix block. To visualize this filter let us consider a hypothetical space group with maximal $\mathbf{k}$-vectors  $A$ and $B$ connected by a line $L$. We take the compatibility relations to be
 \begin{equation}
 A_1\rightarrow L_1\oplus L_2,\quad
 A_2\rightarrow L_1\oplus L_2,\quad
 B_1\rightarrow L_1\oplus L_2,\quad
 B_2\rightarrow L_1\oplus L_2,
 \label{eq:1dequiv}
 \end{equation}
 such that $A_1$, $A_2$, $B_1$ and $B_2$ are irreps of dimension $2$, and $L_1$ and $L_2$ have dimension $1$. We fix the sub-matrix corresponding to the $A-L$ connection as shown in Table~\ref{table:SUB2}.
 \begin{table}[h]
\begin{tabular}{c|cccc}
  & $L_1$ & $L_1$ & $L_2$ & $L_2$\\
 \hline
  ${A_1}$ & 1 & 0 & 1 & 0 \\ 
  ${A_2}$ & 0 & 1 & 0 & 1\end{tabular}
\caption{Fixed sub-matrix block of the $A-L$ connection.}
\label{table:SUB2}
\end{table}

The possible complementary $B-L$ sub-matrices are given in Table~\ref{table:SUB3}.
 \begin{table}[h]
\begin{tabular}{c|cccc||cccc||cccc||cccc}
  & $L_1$ & $L_1$ & $L_2$ & $L_2$ & $L_1$ & $L_1$ & $L_2$ & $L_2$ & $L_1$ & $L_1$ & $L_2$ & $L_2$ & $L_1$ & $L_1$ & $L_2$ & $L_2$ \\
 \hline
  ${B_1}$ & 1 & 0 & 1 & 0  & 0 & 1 & 0 & 1  & 0 & 1 & 1 & 0  & 1 & 0 & 0 & 1 \\ 
  ${B_2}$ & 0 & 1 & 0 & 1  & 1 & 0 & 1 & 0  & 1 & 0 & 0 & 1  & 0 & 1 & 1 & 0
  \end{tabular}
\caption{All possible sub-matrix blocks of the $B-L$ connection.}
\label{table:SUB3}
\end{table}
By examining Fig.~\ref{fig:1}(b), we can see the crossing of irreps corresponding to each of these $B-L$ submatrices. Since crossings of identical irreps can be gapped, it is easy to see that both (I) and (II) in the figure give the same generic band connectivity, and so we only need to consider one of them. 
{To implement this, our algorithm first indexes all valid submatrices for each connection. We then ensure that all submatrices that are fixed in the adjacency matrix are chosen as block matrices, with blocks corresponding to the identity matrix (c.f.~Table~\ref{table:SUB2}). Having done this, we identify connections where two different representations of the little group of a maximal $\mathbf{k}$-vector are connected by an identical pair of non-maximal representations, as along the $A-L-B$ connection from Eq.~(\ref{eq:1dequiv}). It is then straightforward to exclude submatrices that correspond to the ``fake-Weyl'' connections of Fig.~\ref{fig:1}(b).}

\begin{figure}[h]
\centering
    \includegraphics[scale=0.4,angle=0]{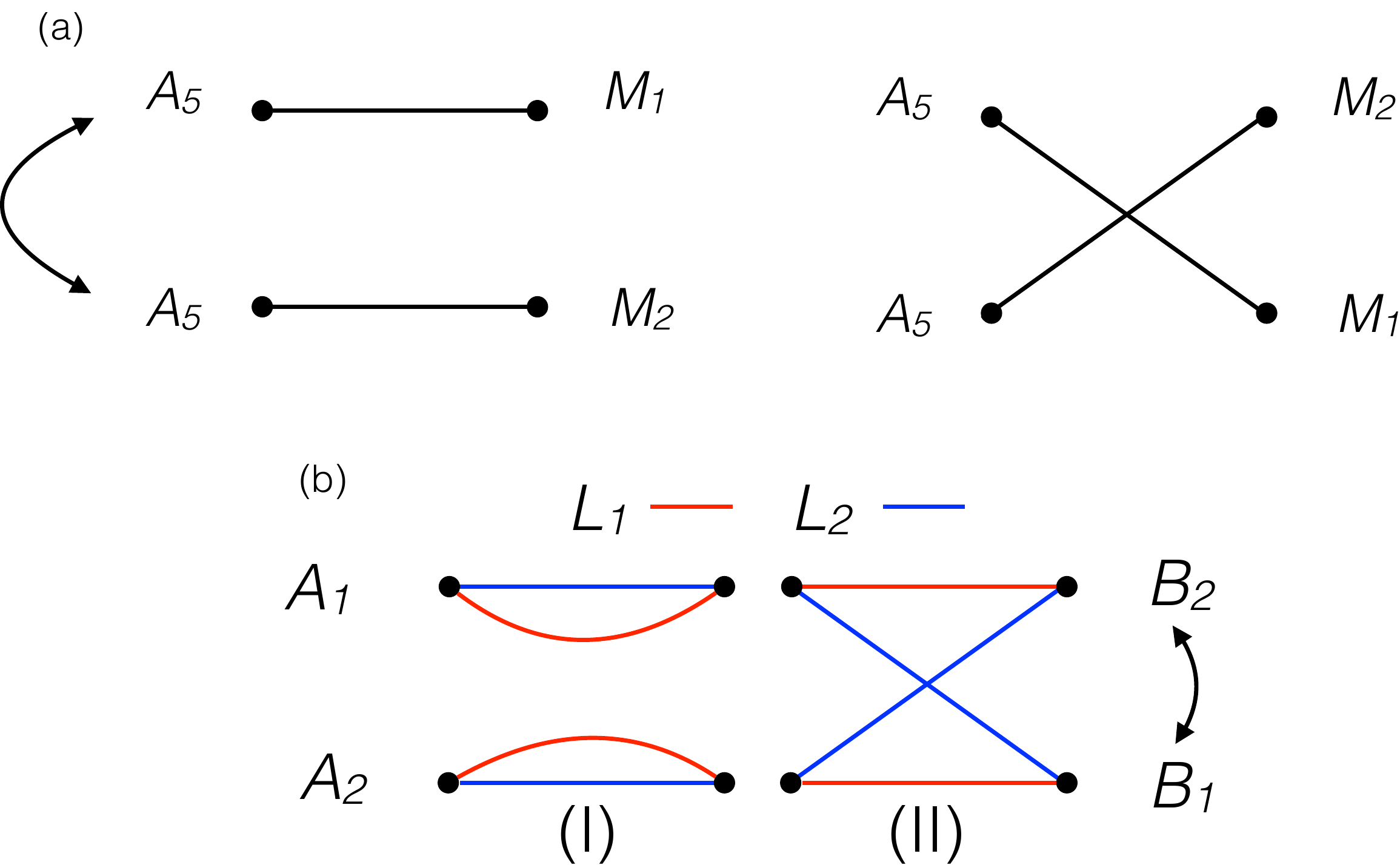}
    \caption{(a) Band structure with two copies of the representation $A_5$ at the point $A$, connected through a line $V$ to distinct representations $M_1$ and $M_2$ at the point $M$. Permuting the representations at $A$ in each case would lead to the same connected elements, since both the left and right subgraphs have the same adjacency matrix. (b) Visual example of fake Weyls. The first two submatrices in Table~\ref{table:SUB3} lead to the connectivity (I), while the last two lead to the connectivity (II). {(The crossing of the two blue lines in (II), which correspond to the same representation, gaps out, and so} we see that these yield the same connectivity).}
\label{fig:1}        
\end{figure}

\item {\it Single-irrep filter}: if one of the $\mathbf{k}$-vectors has only a single little group representation in a given EBR, then all connectivity graphs for this EBR will be fully connected. This follows from the {trivial} observation that all bands must pass through that irrep. Therefore, as long as we are interested only in band connectivity, we do not need to calculate the connectivity graphs for such an EBR. {For example, with TR symmetry, all spin-$1/2$ (double-valued) band representations with 8 bands in space group P$4/ncc$ ($130$) are connected in this way, through an eight-fold degeneracy at the $A$ point\cite{Wieder2016,Bradlyn2016}}.

\end{itemize}

{With these filters in-hand, the gargantuan task of computing all valid connectivity graphs becomes algorithmically tractable\footnote{An additional type of redundancy can occur when only some of the little group representations in a block corresponding to a maximal $\mathbf{k}$-vector are isomorphic (contrast this with the multiple irrep filter, where it was all irreps in a given block). In this case, every permutation of the rows in the full adjacency matrix corresponding to the identical irreps results in an equivalent graph. To account for this, if no submatrices containing this $\mathbf{k}$-vector are fixed by other means, then the first column of the first non-zero submatrix in which this maximal $\mathbf{k}$-vector appears is fixed, and only the rest of the columns will be permuted. However, we have not found it computationally necessary to implement such a filter.}}

\subsection{Disconnected graphs from spectral graph theory}\label{sec:maialaplacian}

We now consider all connectivity graphs consistent with the representation content of an (physically) elementary band representation, constructed using the algorithm described above. Because
the band representation is elementary, we know that the connectivity graphs will have either one connected component,
or will decompose into a set of disconnected components. Furthermore, from Ref.~\onlinecite{NaturePaper}, all connectivity graphs with more than one connected component will then correspond to a possible topological phase. To determine which connectivity graphs are topologically disconnected in this way, we employ the spectral graph partitioning technique described in Sec.~\ref{sec:graphreview}. In particular, the number of zero eigenvalues of the graph Laplacian matrix gives the number of disconnected components of the connectivity graph, and the associated eigenvectors give the little group representations in each component. Let us consider as a first example the space group P4$mm$ ($99$). We focus on the physically elementary band representation (i.~e.~ including TR symmetry) induced from the $\bar{E}$ ($\bar{\Gamma}_{5}$) site-symmetry representation of the $2c$ Wyckoff position EBR. We find that the Laplacian matrices for this band representation are $ 30\times 30$ matrices. We give one example in Table~\ref{table:LAP1}. 
\begin{table} 
\tiny{
\begin{tabular}{cccccccccccccccccccccccccccccc}
 $\bar{A_6}$ & $\bar{A_7}$ & $\bar{\Gamma_6}$ & $\bar{\Gamma_7}$ & $\bar{M_6}$ & $\bar{M_7}$ & $\bar{R_5}$ & $\bar{R_5}$ & $\bar{X_5}$        & $\bar{X_5}$ & $\bar{Z_6}$ & $\bar{Z_7}$ & $\bar{V_6}$ & $\bar{V_7}$ & $\bar{C_3}$ & $\bar{C_3}$ & $\bar{C_4}$ & $\bar{C_4}$ & $\bar{T_3}$ & $\bar{T_3}$ & $\bar{T_4}$& $\bar{T_4}$& $\bar{\Delta_6}$& $\bar{\Delta_7}$& $\bar{B_3}$& $\bar{B_3}$& $\bar{B_4}$& $\bar{B_4}$& $\bar{W_5}$& $\bar{W_5}$ \\
 \hline
6 &0 & 0 & 0 & 0 &0 &0  &0 &0 &0 &0 &0 &-2&0 &-1& 0 &-1&0 &-1&0 &-1&0 &0 &0 &0 &0 &0 &0 &0 &0 \\ 
0 &6 & 0 & 0 & 0 &0 &0  &0 &0 &0 &0 &0 &0 &-2& 0& -1&0 &-1&0 &-1&0 &-1&0 &0 &0 &0 &0 &0 &0 &0 \\ 
0 &0 & 6 & 0 & 0 &0 &0  &0 &0 &0 &0 &0 &0 &0 &-1& 0 &-1&0 &0 &0 &0 &0 &-2&0 &-1&0 &-1&0 &0 &0 \\ 
0 &0 & 0 & 6 & 0 &0 &0  &0 &0 &0 &0 &0 &0 &0 &0 & -1&0 &-1&0 &0 &0 &0 &0 &-2&0 &-1&0 &-1&0 &0 \\ 
0 &0 & 0 & 0 & 2 &0 &0  &0 &0 &0 &0 &0 &-2&0 &0 & 0 &0 &0 &0 &0 &0 &0 &0 &0 &0 &0 &0 &0 &0 &0 \\ 
0 &0 & 0 & 0 & 0 &2 &0  &0 &0 &0 &0 &0 &0 &-2&0 & 0 &0 &0 &0 &0 &0 &0 &0 &0 &0 &0 &0 &0 &0 &0 \\ 
0 &0 & 0 & 0 & 0 &0 &6  &0 &0 &0 &0 &0 &0 &0 &0 & 0 &0 &0 &-1&0 &-1&0 &0 &0 &-1&0 &-1&0 &0 &-2\\
0 &0 & 0 & 0 & 0 &0 &0  &6 &0 &0 &0 &0 &0 &0 &0 & 0 &0 &0 &0 &-1&0 &-1&0 &0 &0 &-1&0 &-1&-2&0 \\
0 &0 & 0 & 0 & 0 &0 &0  &0 &2 &0 &0 &0 &0 &0 &0 & 0 &0 &0 &0 &0 &0 &0 &0 &0 &0 &0 &0 &0 &0 &-2\\
0 &0 & 0 & 0 & 0 &0 &0  &0 &0 &2 &0 &0 &0 &0 &0 & 0 &0 &0 &0 &0 &0 &0 &0 &0 &0 &0 &0 &0 &-2&0 \\ 
0 &0 & 0 & 0 & 0 &0 &0  &0 &0 &0 &2 &0 &0 &0 &0 & 0 &0 &0 &0 &0 &0 &0 &-2&0 &0 &0 &0 &0 &0 &0 \\ 
0 &0 & 0 & 0 & 0 &0 &0  &0 &0 &0 &0 &2 &0 &0 &0 & 0 &0 &0 &0 &0 &0 &0 &0 &-2&0 &0 &0 &0 &0 &0 \\ 
-2&0 & 0 & 0 & -2&0 &0  &0 &0 &0 &0 &0 &4 &0 &0 & 0 &0 &0 &0 &0 &0 &0 &0 &0 &0 &0 &0 &0 &0 &0 \\ 
0 &-2& 0 & 0 & 0 &-2&0  &0 &0 &0 &0 &0 &0 &4 &0 & 0 &0 &0 &0 &0 &0 &0 &0 &0 &0 &0 &0 &0 &0 &0 \\ 
-1&0 & -1& 0 & 0 &0 &0  &0 &0 &0 &0 &0 &0 &0 &2 & 0 &0 &0 &0 &0 &0 &0 &0 &0 &0 &0 &0 &0 &0 &0 \\ 
0 &-1& 0 & -1& 0 &0 &0  &0 &0 &0 &0 &0 &0 &0 &0 & 2 &0 &0 &0 &0 &0 &0 &0 &0 &0 &0 &0 &0 &0 &0 \\ 
-1&0 & -1& 0 & 0 &0 &0  &0 &0 &0 &0 &0 &0 &0 &0 & 0 &2 &0 &0 &0 &0 &0 &0 &0 &0 &0 &0 &0 &0 &0 \\ 
0 &-1& 0 & -1& 0 &0 &0  &0 &0 &0 &0 &0 &0 &0 &0 & 0 &0 &2 &0 &0 &0 &0 &0 &0 &0 &0 &0 &0 &0 &0 \\ 
-1&0 & 0 & 0 & 0 &0 &-1 &0 &0 &0 &0 &0 &0 &0 &0 & 0 &0 &0 &2 &0 &0 &0 &0 &0 &0 &0 &0 &0 &0 &0 \\ 
0 &-1& 0 & 0 & 0 &0 &0  &-1&0 &0 &0 &0 &0 &0 &0 & 0 &0 &0 &0 &2 &0 &0 &0 &0 &0 &0 &0 &0 &0 &0 \\ 
-1&0 & 0 & 0 & 0 &0 &-1 &0 &0 &0 &0 &0 &0 &0 &0 & 0 &0 &0 &0 &0 &2 &0 &0 &0 &0 &0 &0 &0 &0 &0 \\ 
0 &-1& 0 & 0 & 0 &0 &0  &-1&0 &0 &0 &0 &0 &0 &0 & 0 &0 &0 &0 &0 &0 &2 &0 &0 &0 &0 &0 &0 &0 &0 \\ 
0 &0 & -2& 0 & 0 &0 &0  &0 &0 &0 &-2&0 &0 &0 &0 & 0 &0 &0 &0 &0 &0 &0 &4 &0 &0 &0 &0 &0 &0 &0 \\ 
0 &0 & 0 & -2& 0 &0 &0  &0 &0 &0 &0 &-2&0 &0 &0 & 0 &0 &0 &0 &0 &0 &0 &0 &4 &0 &0 &0 &0 &0 &0 \\ 
0 &0 & -1& 0 & 0 &0 &-1 &0 &0 &0 &0 &0 &0 &0 &0 & 0 &0 &0 &0 &0 &0 &0 &0 &0 &2 &0 &0 &0 &0 &0 \\ 
0 &0 & 0 & -1& 0 &0 &0  &-1&0 &0 &0 &0 &0 &0 &0 & 0 &0 &0 &0 &0 &0 &0 &0 &0 &0 &2 &0 &0 &0 &0 \\ 
0 &0 & -1& 0 & 0 &0 &-1 &0 &0 &0 &0 &0 &0 &0 &0 & 0 &0 &0 &0 &0 &0 &0 &0 &0 &0 &0 &2 &0 &0 &0 \\ 
0 &0 & 0 & -1& 0 &0 &0  &-1&0 &0 &0 &0 &0 &0 &0 & 0 &0 &0 &0 &0 &0 &0 &0 &0 &0 &0 &0 &2 &0 &0 \\ 
0 &0 & 0 & 0 & 0 &0 &0  &-2&0 &-2&0 &0 &0 &0 &0 & 0 &0 &0 &0 &0 &0 &0 &0 &0 &0 &0 &0 &0 &4 &0 \\ 
0 &0 & 0 & 0 & 0 &0 &-2 &0 &-2&0 &0 &0 &0 &0 &0 & 0 &0 &0 &0 &0 &0 &0 &0 &0 &0 &0 &0 &0 &0 &4
\end{tabular}
}
\caption{Schematic plot of the Laplacian matrix for the physically elementary band representation of space group P4$mm$ ($99$) induced from the $\bar{\Gamma}_{5}$ representation of the stabilizer group of the $2c$ Wyckoff position.}
\label{table:LAP1}
\end{table}
\begin{sidewaystable}
\tiny
\miniscule{
\begin{tabular}{c|cccccccccccccccccccccccccccccccccccccccccccccccc}
&$\bar{S_2}$& $\bar{S_2}$& $\bar{S_3}$& $\bar{S_3}$& $\bar{S_4}$& $\bar{S_4}$& $\bar{S_5}$& $\bar{S_5}$& $\bar{V_6}$&$\bar{V_6}$ &$\bar{V_7}$& $\bar{V_7}$& $\bar{T_5}$ & $\bar{T_5}$ & $\bar{T_5}$ & $\bar{T_5}$  & $\bar{\Delta_5}$ & $\bar{\Delta_5}$ & $\bar{\Delta_5}$ & $\bar{\Delta_5}$ & $\bar{\Lambda_6}$ & $\bar{\Lambda_6}$ & $\bar{\Lambda_7}$ & $\bar{\Lambda_7}$ & $\bar{\Sigma_5}$ & $\bar{\Sigma_5}$ & $\bar{\Sigma_5}$& $\bar{\Sigma_5}$  & $\bar{Y_2}$& $\bar{Y_2}$ & $\bar{Y_3}$& $\bar{Y_3}$& $\bar{Y_4}$& $\bar{Y_4}$& $\bar{Y_5}$& $\bar{Y_5}$& $\bar{U_2}$& $\bar{U_2}$& $\bar{U_3}$& $\bar{U_3}$& $\bar{U_4}$& $\bar{U_4}$& $\bar{U_5}$& $\bar{U_5}$ & $\bar{W_5}$& $\bar{W_5}$& $\bar{W_5}$& $\bar{W_5}$\\
\hline
$\bar{A_5}$&0 & -1 & 0 & -1 & 0 & -1 & 0 & -1 & -1 & 0 & -1 & 0 & 0 & 0 & -2 & -2 &  0 & 0 & 0 & 0 &  0 & 0 & 0 & 0 & 0 & 0 & 0 & 0 & 0 & 0 & 0 & 0 & 0 & 0 & 0 & 0 & 0 & 0 & 0 & 0 & 0 & 0 & 0 & 0 & 0 & 0 & 0 & 0 \\
$\bar{A_5}$&-1 & 0 & -1 & 0 & -1 & 0 & -1 & 0 & 0 & -1 &  0 & -1 & -2 & -2 & 0 & 0 & 0 & 0 & 0 & 0 &  0 & 0 & 0 & 0 & 0 & 0 & 0 & 0 & 0 & 0 & 0 & 0 & 0 & 0 & 0 & 0 & 0 & 0 & 0 & 0 & 0 & 0 & 0 & 0 & 0 & 0 & 0 & 0 \\
$\bar{\Gamma_6}$&0 & 0 & 0 & 0 & 0 & 0 & 0 & 0 & 0 & 0 & 0 & 0 & 0 & 0 & 0 & 0 & 0 & 0 & 0 & -2 &  0 & 0 & 0 & -2 & 0 & -2 & 0 & 0 & 0 & 0 & 0 & 0 & 0 & 0 & 0 & 0 & 0 & 0 & 0 & 0 & 0 & 0 & 0 & 0 & 0 & 0 & 0 & 0 \\
$\bar{\Gamma_6}$&0 & 0 & 0 & 0 & 0 & 0 & 0 & 0 & 0 & 0 & 0 & 0 & 0 & 0 & 0 & 0 & 0 & 0 & -2 & 0 &  0 & 0 & -2 & 0 & -2 & 0 & 0 & 0 & 0 & 0 & 0 & 0 & 0 & 0 & 0 & 0 & 0 & 0 & 0 & 0 & 0 & 0 & 0 & 0 & 0 & 0 & 0 & 0 \\
$\bar{\Gamma_7}$&0 & 0 & 0 & 0 & 0 & 0 & 0 & 0 & 0 & 0 & 0 & 0 & 0 & 0 & 0 & 0 &  0 & -2 & 0 & 0 &  0 & -2 & 0 & 0 & 0 & 0 & 0 & -2 & 0 & 0 & 0 & 0 & 0 & 0 & 0 & 0 & 0 & 0 & 0 & 0 & 0 & 0 & 0 & 0 & 0 & 0 & 0 & 0 \\
$\bar{\Gamma_7}$&0 & 0 & 0 & 0 & 0 & 0 & 0 & 0 & 0 & 0 & 0 & 0 & 0 & 0 & 0 & 0 &  -2 & 0 & 0 & 0 &  -2 & 0 & 0 & 0 & 0 & 0 & -2 & 0 & 0 & 0 & 0 & 0 & 0 & 0 & 0 & 0 & 0 & 0 & 0 & 0 & 0 & 0 & 0 & 0 & 0 & 0 & 0 & 0 \\
$\bar{M_5}$&0 & 0 & 0 & 0 & 0 & 0 & 0 & 0 & -1 & 0 & -1 & 0 & 0 & 0 & 0 & 0 &  0 & 0 & 0 & 0 &  0 & 0 & -2 & -2 & 0 & 0 & 0 & 0 & 0 & -1 & 0 & -1 & 0 & -1 & 0 & -1 & 0 & 0 & 0 & 0 & 0 & 0 & 0 & 0 & 0 & 0 & 0 & 0 \\
$\bar{M_5}$&0 & 0 & 0 & 0 & 0 & 0 & 0 & 0 & 0 & -1 & 0 & -1 & 0 & 0 & 0 & 0 &  0 & 0 & 0 & 0 &  -2 & -2 &  0 & 0 & 0 & 0 & 0 & 0 & -1 & 0 & -1 & 0 & -1 & 0 & -1 & 0 & 0 & 0 & 0 & 0 & 0 & 0 & 0 & 0 & 0 & 0 & 0 & 0 \\
$\bar{R_3}$&0 & 0 & 0 & 0 & 0 & 0 & 0 & 0 & 0 & 0 & 0 & 0 & 0 & 0 & 0 & -2 &  0 & 0 & 0 & 0 & 0 & 0 & 0 & 0 & 0 & 0 & 0 & 0 & 0 & 0 & 0 & 0 & 0 & 0 & 0 & 0 & 0 & -1 & 0 & 0 & 0 & -1 & 0 & 0 & 0 & 0 & 0 & -2 \\
$\bar{R_3}$&0 & 0 & 0 & 0 & 0 & 0 & 0 & 0 & 0 & 0 & 0 & 0 & 0 & 0 & -2 & 0 &  0 & 0 & 0 & 0 &  0 & 0 & 0 & 0 & 0 & 0 & 0 & 0 & 0 & 0 & 0 & 0 & 0 & 0 & 0 & 0 & -1 & 0 & 0 & 0 & -1 & 0 & 0 & 0 & 0 & 0 & -2 & 0 \\
$\bar{R_4}$&0 & 0 & 0 & 0 & 0 & 0 & 0 & 0 & 0 & 0 & 0 & 0 & 0 & -2 & 0 & 0 &  0 & 0 & 0 & 0 &  0 & 0 & 0 & 0 & 0 & 0 & 0 & 0 & 0 & 0 & 0 & 0 & 0 & 0 & 0 & 0 & 0 & 0 & 0 & -1 & 0 & 0 & 0 & -1 & 0 & -2 & 0 & 0 \\
$\bar{R_4}$&0 & 0 & 0 & 0 & 0 & 0 & 0 & 0 & 0 & 0 & 0 & 0 & -2 & 0 & 0 & 0 &  0 & 0 & 0 & 0 &  0 & 0 & 0 & 0 & 0 & 0 & 0 & 0 & 0 & 0 & 0 & 0 & 0 & 0 & 0 & 0 & 0 & 0 & -1 & 0 & 0 & 0 & -1 & 0 & -2 & 0 & 0 & 0 \\
$\bar{X_3}$&0 & 0 & 0 & 0 & 0 & 0 & 0 & 0 & 0 & 0 & 0 & 0 & 0 & 0 & 0 & 0 &  0 & 0 & 0 & -2 &  0 & 0 & 0 & 0 & 0 & 0 & 0 & 0 & 0 & -1 & 0 & 0 & 0 & -1 & 0 & 0 & 0 & 0 & 0 & 0 & 0 & 0 & 0 & 0 & 0 & 0 & 0 & -2 \\
$\bar{X_3}$&0 & 0 & 0 & 0 & 0 & 0 & 0 & 0 & 0 & 0 & 0 & 0 & 0 & 0 & 0 & 0 &  0 & 0 & -2 & 0 &  0 & 0 & 0 & 0 & 0 & 0 & 0 & 0 & -1 & 0 & 0 & 0 & -1 & 0 & 0 & 0 & 0 & 0 & 0 & 0 & 0 & 0 & 0 & 0 & 0 & 0 & -2 & 0 \\
$\bar{X_4}$&0 & 0 & 0 & 0 & 0 & 0 & 0 & 0 & 0 & 0 & 0 & 0 & 0 & 0 & 0 & 0 &  0 & -2 & 0 & 0 &  0 & 0 & 0 & 0 & 0 & 0 & 0 & 0 & 0 & 0 & 0 & -1 &  0 & 0 & 0 & -1 & 0 & 0 & 0 & 0 & 0 & 0 &        0 & 0 & 0 & -2 & 0 & 0 \\
$\bar{X_4}$&0 & 0 & 0 & 0 & 0 & 0 & 0 & 0 & 0 & 0 & 0 & 0 & 0 & 0 & 0 & 0 &  -2 & 0 & 0 & 0 &  0 & 0 & 0 & 0 & 0 & 0 & 0 & 0 & 0 & 0 & -1 & 0 &        0 & 0 & -1 & 0 & 0 & 0 & 0 & 0 & 0 & 0 &        0 & 0  & -2 & 0 & 0 & 0 \\
$\bar{Z_5}$&0 & -1 & 0 & 0 & 0 & 0 & 0 & -1 & 0 & 0 & 0 & 0 & 0 & 0 & 0 & 0 &  0 & 0 & 0 & 0 &  0 & 0 & 0 & 0 & 0 & -2 & 0 & 0 & 0 & 0 & 0 & 0 &        0 & 0 & 0 & 0 & 0 & 0 & 0 & -1 & 0 & -1 &       0 & 0 & 0 & 0 & 0 & 0 \\
$\bar{Z_6}$&-1 & 0 & 0 & 0 & 0 & 0 & -1 & 0 & 0 & 0 & 0 & 0 & 0 & 0 & 0 & 0 &  0 & 0 & 0 & 0 &  0 & 0 & 0 & 0 & 0 & 0 & 0 & -2 & 0 & 0 & 0 & 0 &        0 & 0 & 0 & 0 & 0 & 0 & -1 & 0 & -1 & 0 &       0 & 0 & 0 & 0 & 0 & 0 \\
$\bar{Z_7}$&0 & 0 & 0 & -1 & 0 & -1 & 0 & 0 & 0 & 0 & 0 & 0 & 0 & 0 & 0 & 0 &  0 & 0 & 0 & 0 &  0 & 0 & 0 & 0 & -2 & 0 & 0 & 0 & 0 & 0 & 0 & 0 &        0 & 0 & 0 & 0 & 0 & -1 & 0 & 0 & 0 & 0 &        0 & -1 & 0 & 0 & 0 & 0 \\
$\bar{Z_8}$&0 & 0 & -1 & 0 & -1 & 0 & 0 & 0 & 0 & 0 & 0 & 0 & 0 & 0 & 0 & 0 &  0 & 0 & 0 & 0 &  0 & 0 &        0 & 0 & 0 & 0 & -2 & 0 & 0 & 0 & 0 & 0 &        0 & 0 & 0 & 0 & -1 & 0 & 0 & 0 & 0 & 0 &        -1 & 0 & 0 & 0 & 0 & 0  
\end{tabular}
}
\caption{Schematic plot of part of the Laplacian matrix of P4$/ncc$ 8d $\bar{\Gamma}_{3}$ EBR.}
\label{table:LAP2}
\end{sidewaystable}

We used the {\it NumPy} package to diagonalize the matrix. In the example of Table~\ref{table:LAP1}, we find that there are two zero eigenvectors. We thus have two $30$-dimensional eigenvectors with $2$ different values among their 30 components, this means the EBR can be disconnected in 2 different ways. Because there are multiple null eigenvectors, numerical diagonalization will generically return an arbitrary linear combination of them, rather than giving them in a basis where all entries are either $1$ or $0$ {(any linear combination of degenerate eigenvectors is also an eigenvector)}. To find this basis, we first form the (non-square) matrix whose rows are the zero eigenvectors. We then perform Gauss-Jordan elimination on this matrix, yielding a matrix whose entries are all either zero or one. The rows of this matrix are then the eigenvectors in the appropriate basis. We can then read off the irreps in each connected component. To see how this works in detail, let us return to the Laplacian matrix in Table~\ref{table:LAP1}. The numerically determined eigenvectors are
\begin{equation}
\begin{array}{lll}
{\bf{v}_1} = (0.258, -0.0098, 0.258, -0.0098, 0.258, -0.0098, 0.258, -0.0098, 0.258, -0.0098, 0.258, -0.0098, \\
0.258, -0.0098, 0.258, -0.0098, 0.258, -0.0098, 0.258, -0.0098, 0.258, -0.0098, 0.258, -0.0098, 0.258,\\
 -0.0098, 0.258, -0.0098, 0.258, -0.0098) \\
 \\
{\bf{v}_2} = (0.0098, 0.258, 0.0098, 0.258, 0.0098, 0.258, 0.0098, 0.258, 0.0098, 0.258, 0.0098, 0.258, \\
0.0098, 0.258, 0.0098, 0.258, 0.0098, 0.258, 0.0098, 0.258, 0.0098, 0.258, 0.0098, 0.258, 0.0098,\\
 0.258, 0.0098, 0.258, 0.0098, 0.258).
\end{array}
\end{equation}
Forming the matrix 
\begin{equation}
\left[\begin{array}{c}
\mathbf{v}_1 \\
\mathbf{v}_2
\end{array}\right]
\end{equation}
 and performing Gaussian elimination yields
\begin{equation}
\left[\begin{array}{c}
\mathbf{v}_1 \\
\mathbf{v}_2
\end{array}\right]\rightarrow\left[
\begin{array}{cccccccccccccccccccccccccccccc}
1 & 0 & 1 & 0 & 1 & 0 & 1 & 0 & 1 & 0 & 1 & 0 & 1 & 0 & 1 & 0 & 1 & 0 & 1 & 0 & 1 & 0 & 1 & 0 & 1 & 0 & 1 & 0 & 1 & 0 \\
0 & 1 & 0 & 1 & 0 & 1 & 0 & 1 & 0 & 1 & 0 & 1 & 0 & 1 & 0 & 1 & 0 & 1 & 0 & 1 & 0 & 1 & 0 & 1 & 0 & 1 & 0 & 1 & 0 & 1
\end{array}\right].
\end{equation}
 Comparing the rows of this matrix with the ordering of representations in the Laplacian matrix shown in Table~\ref{table:LAP1}, we find for the two connected components of this connectivity graph
\begin{equation}
\bar{A_6} - \bar{\Gamma_6} - \bar{M_6} - \bar{R_5} - \bar{X_5} - \bar{Z_6} - \bar{V_6} - \bar{C_3} -\bar{C_4} -\bar{T_3} -\bar{T_4} -\bar{\Lambda_6} -\bar{B_3} -\bar{B_4} -\bar{W_5},
\end{equation}
and
\begin{equation}
\bar{A_7} -\bar{\Gamma_7} - \bar{M_7} -\bar{R_5} -\bar{X_5} -\bar{Z_7} -\bar{V_7} -\bar{C_3} -\bar{C_4} -\bar{T_3} -\bar{T_4} -\bar{\Lambda_7} -\bar{B_3} -\bar{B_4} -\bar{W_5},
\end{equation}
and so the band structure corresponding to this connectivity graph has two topologically disconnected groups of bands. This procedure based on Gauss-Jordan elimination generalizes straightforwardly to Laplacian matrices with more than two zero eigenvalues.
%

Let us now return to our previous example, the band representation of space group $P4/ncc$ ($130$) induced from the $\bar{\Gamma}_{3}$ representation of the stabilizer group of the $8d$ Wyckoff position. We find that there are approximately $\sim15\times10^{6}$ valid $ 80\times 80$ Laplacian matrices. The diagonal entries of each of these correspond to the degree matrix, and are all given by
\begin{equation}
\begin{array}{lll}
D = (16,16,8,8,8,8,16,16,8,8,8,8,8,8,8,8,8,8,8,8,2,2,2,2,2,2,2,2,4,4,4,4,4,4,4,4,4,4,4,4, \\
4,4,4,4,4,4,4,4,4,4,4,4,4,4,4,4,2,2,2,2,2,2,2,2,2,2,2,2,2,2,2,2,4,4,4,4,4,4,4,4).
\end{array}
\end{equation}
In Table~\ref{table:LAP2} we give one of these Laplacian matrices. In the interest of space, we show only the rows above the main diagonal that are not trivially zero. Diagonalizing all valid Laplacian matrices, we find that this band representation can be disconnected in two inequivalent ways. In the first, the two disconnected components contain the representations
\begin{equation}
\bar{A_5}-\bar{\Gamma_8}-\bar{\Gamma_8}-\bar{M_5}-\bar{R_3}-\bar{R_4}-\bar{X_3}-\bar{X_4}-\bar{Z_5}-\bar{Z_7}\label{eq:decomp1}
\end{equation}
and
\begin{equation}
\bar{A_5}-\bar{\Gamma_9}-\bar{\Gamma_9}-\bar{M_5}-\bar{R_3}-\bar{R_4}-\bar{X_3}-\bar{X_4}-\bar{Z_6}-\bar{Z_8}\label{eq:decomp2}.
\end{equation}  
Alternatively, we find that the decomposition into components
\begin{equation}
\bar{A_5}-\bar{\Gamma_8}-\bar{\Gamma_9}-\bar{M_5}-\bar{R_3}-\bar{R_4}-\bar{X_3}-\bar{X_4}-\bar{Z_5}-\bar{Z_8}\label{eq:decomp3}
\end{equation}
and
\begin{equation}
\bar{A_5}-\bar{\Gamma_8}-\bar{\Gamma_9}-\bar{M_5}-\bar{R_3}-\bar{R_4}-\bar{X_3}-\bar{X_4}-\bar{Z_6}-\bar{Z_7}\label{eq:decomp4}
\end{equation}
is also possible.

Since we construct the full connectivity graph for each band representation, we can also deduce the existence of symmetry-enforced band crossings in semi-metallic band structures. {For example, we can see from the decomposition Eqs.~(\ref{eq:decomp1}--\ref{eq:decomp4}) that at $1/4$ filling (which can be achieved, for instance, by charge transfer) the band representation in space group P$4/ncc$ ($130$) induced from the $\bar{\Gamma}_3$ representation of the site symmetry group of the $8d$ Wyckoff position realize a semimetal, with nodal Fermi surfaces at the $A$ and $M$ points.} For cases where the information about semimetals is not needed, we will now develop an algorithm in Sec.~\ref{sec:luisgraph} which is more efficient at determining the allowed decompositions of a connectivity graph. This algorithm determines disconnected components without explicitly constructing the full graph in the cases where it is connected.

\subsection{Disconnected graphs via {fast} search}\label{sec:luisgraph}

We now present a second algorithm, which directly constructs disconnected connectivity graphs. While it is based almost {purely on} combinatorial analysis, it is related in spirit to Prim's algorithm for finding minimal spanning forests\cite{prim1957shortest}, in that it ``grows'' a disconnected connectivity graph starting from a set of seed nodes. Because the algorithm terminates once a solution is found, it has a much faster average runtime than the direct method of Sec.~\ref{sec:maiagraph}. We start with the (previously calculated\cite{grouptheory}) multiplicities of all the irreps of the little group of every maximal $\mathbf{k}$-vector in the decomposition of a band representation, the independent set of paths between them, and all the compatibility relations along these paths. We then attempt to partition the set of irreps at every maximal $\mathbf{k}$-vector into subsets in such a way that a connectivity graph can be constructed with each subset corresponding to a distinct disconnected subgraph. These subgraphs can, in principle, be further disconnected into smaller connectivity subgraphs. The purpose of our algorithm is to determine all the possible ways to decompose an elementary band representation into disconnected subgraphs that are further indecomposable. 

We call each of these indecomposable subgraphs a \emph{branch} of the connectivity graph of the band representation. Every distinct decomposition into branches represents a valid connectivity graph. Throughout, we shall describe the whole process with our example of space group $P4/ncc$ ($130$); {the general formalism then becomes obvious}. We will focus on the elementary band representation induced from the $A_g$ (in the notation of Ref.~\onlinecite{mulliken} or $\Gamma_1^+$ in the notation of Ref.~\onlinecite{koster}) irrep of the site symmetry group of the Wyckoff position $8d$ {(with representative coordinates $(0,0,0)$ in the unit cell)}, isomorphic to the point group $\bar{1}$, {generated by inversion}. Table \ref{table:subducedirreps} gives the subduced irreps of this band representation into every maximal $\mathbf{k}^M$ (see Table \ref{table:maximalkvecs}). For simplicity, we will not consider TR symmetry in the example.

At each maximal $\mathbf{k}$-vector $\mathbf{k}^M$, we calculate two parameters $N(\mathbf{k}^M)$ and $\Omega(\mathbf{k}^M)$ defined as follows. If the elementary band representation subduces into the irreps $\rho_1,\rho_2,\ldots...$ of the little group of $\mathbf{k}^M$ with multiplicities $n_1,n_2,\ldots$, then
\begin{equation}
N(\mathbf{k}^M)=\sum_in_i\hspace{1cm}\textrm{and}\hspace{1cm}\Omega(\mathbf{k}^M)=\frac{N!}{n_1!n_2!\ldots}
\end{equation}
are the total number of irreps at $\mathbf{k}^M$ and the number of distinguishable ways to order these irreps (according to the energy associated to each irrep, for example). {Next, we impose an ordering on the maximal $\mathbf{k}$-vectors $\{\mathbf{k}^M_1,\mathbf{k}^M_2,\dots\}$ of the space group. This ordering is defined by the rules
\begin{align}
N(\mathbf{k}^M_i)&\leq N(\mathbf{k}^M_j),\;\mathrm{if}\; i<j \\
\Omega(\mathbf{k}^M_i)&\leq\Omega(\mathbf{k}^M_j),\;\mathrm{if}\;i<j\;\mathrm{and}\;N(\mathbf{k}^M_i)=N(\mathbf{k}^M_j)
\end{align}
If two or more $\mathbf{k}$-vectors have the same values of both $N$ and $\Omega$, then we order them in an arbitrary way. 
We can thus refer to an elementary band representation by the label of the first vector $\mathbf{k}^M_1$ in this ordering. }
In the example of the elementary band representation of Table \ref{table:subducedirreps}, the maximal $\mathbf{k}$-vectors are ordered according to these rules. For this elementary band representation, {$\mathbf{k}^M_1\equiv R$, and we have $N(R)=4$ and $\Omega(R)=6$. This reflects the fact that the $R$ point there are two distinct irreps $R_1$ and $R_2$, each with multiplicity two; these four irreps can be ordered in $6$ distinguishable ways. }


Next, we calculate the compatibility relations between every irrep at every maximal $\mathbf{k}^M_i$ and the irreps at the non-redundant lines or planes connected to it, as per Sec.~\ref{subsec:independentpaths}. For our example, Table \ref{table:subducedcomprels} gives these compatibility relations for the elementary band representation of Table \ref{table:subducedirreps}. The first column {of Table~\ref{table:subducedcomprels}} gives the maximal $\mathbf{k}$-vectors $\mathbf{k}^{M_i}$, the second column lists, for each $\mathbf{k}^{M_i}$, the irreps of its little group obtained by subduction of the elementary band representation {this information is also given in Table~\ref{table:subducedirreps})}. The third and fourth columns give the dimension and the multiplicity of the irrep. The next columns give the compatibility relations for each irrep of each $\mathbf{k}^{M_i}$ along the paths that connect this point with the point shown at the first row of the {given} column. Only the independent paths, according to the discussion of subsection \ref{subsec:independentpaths} are considered; we denote by "-" those paths that contain redundant information, and which can be omitted in the analysis.

\begin{table}
\begin{tabular}{ll|l}
$\mathbf{k}$-vec&coordinates&subduced irreps\\
\hline
$R$&$(0,1/2,1/2)$& 2 $R_1(2) \oplus$ 2 $R_2(2)$\\
$X$&$(0,1/2,0)$& 2 $X_1(2) \oplus$ 2 $X_2(2)$\\
$M$&$(1/2,1/2,0)$&$M_1(2) \oplus M_2(2) \oplus M_3(2) \oplus M_4(2)$\\
$A$&$(1/2,1/2,1/2)$&$A_1(2) \oplus A_2(2) \oplus A_3(2) \oplus A_4(2)$\\
$Z$&$(0,0,1/2)$&$Z_1(2) \oplus Z_2(2) \oplus Z_3(2) \oplus Z_4(2)$\\
$\Gamma$&$(0,0,0)$&$\Gamma_1^+(1) \oplus \Gamma_2^+(1) \oplus \Gamma_3^+(1) \oplus$  $\Gamma_4^+(1) \oplus$ 2 $\Gamma_5^+(2)$
\end{tabular}
\caption{Decomposition of the elementary band representation induced from the Wyckoff position $8d$ of the space group P$4/ncc$ ($130$), with site-symmetry group isomorphic to $\bar{1}$, and irrep $A_g$ ($\Gamma_1^+$) into irreps of the litte group at each maximal $\mathbf{k}$-vector. The number in parenthesis shows the dimension of the representation. The number of bands in the elementary band representation is 8.}
\label{table:subducedirreps}
\end{table}

\begin{table}
\begin{tabular}{llll|c|c|c|c|c|c}
&irrep&dim.&mult&$R:(0,1/2,1/2)$&$X:(0,1/2,0)$&$M:(1/2,1/2,0)$&$A:(1/2,1/2,1/2)$&$Z:(0,0,1/2)$&$\Gamma:(0,0,0)$\\
\hline
$R:(0,1/2,1/2)$&$R_1$&2&2&-&$W_1$,$W_4$&-&$T_1$,$T_3$&2 $U_1$&-\\
&$R_2$&2&2&-&$W_2$,$W_3$&-&$T_2$,$T_4$&2 $U_1$&-\\
\hline
$X:(0,1/2,0)$&$X_1$&2&2&$W_1$,$W_3$&-&$Y_1$&-&-&$\Delta_1$,$\Delta_3$\\
&$X_2$&2&2&$W_2$,$W_4$&-&$Y_1$&-&-&$\Delta_2$,$\Delta_4$\\
\hline
$M:(1/2,1/2,0)$&$M_1$&2&1&-&$Y_1$&-&$V_5$&-&$\Sigma_2$,$\Sigma_3$\\
&$M_2$&2&1&-&$Y_1$&-&$V_5$&-&$\Sigma_1$,$\Sigma_4$\\
&$M_3$&2&1&-&$Y_1$&-&$V_1$,$V_3$&-&$\Sigma_1$,$\Sigma_3$\\
&$M_4$&2&1&-&$Y_1$&-&$V_2$,$V_4$&-&$\Sigma_2$,$\Sigma_4$\\
\hline
$A:(1/2,1/2,1/2)$&$A_1$&2&1&$T_1$,$T_2$&-&$V_5$&-&$S_1$&-\\
&$A_2$&2&1&$T_3$,$T_4$&-&$V_5$&-&$S_1$&-\\
&$A_3$&2&1&$T_2$,$T_4$&-&$V_1$,$V_2$&-&$S_1$&-\\
&$A_4$&2&1&$T_1$,$T_3$&-&$V_3$,$V_4$&-&$S_1$&-\\
\hline
$Z:(0,0,1/2)$&$Z_1$&2&1&2 $U_1$&-&-&$S_1$&-&$\Lambda_2$,$\Lambda_3$\\
&$Z_2$&2&1&2 $U_1$&-&-&$S_1$&-&$\Lambda_1$,$\Lambda_4$\\
&$Z_3$&2&1&2 $U_1$&-&-&$S_1$&-&$\Lambda_5$\\
&$Z_4$&2&1&2 $U_1$&-&-&$S_1$&-&$\Lambda_5$\\
\hline
$\Gamma:(0,0,0)$&$\Gamma_1^+$&1&1&-&$\Delta_1$&$\Sigma_1$&-&$\Lambda_1$&-\\
&$\Gamma_2^+$&1&1&-&$\Delta_1$&$\Sigma_2$&-&$\Lambda_2$&-\\
&$\Gamma_3^+$&1&1&-&$\Delta_4$&$\Sigma_2$&-&$\Lambda_4$&-\\
&$\Gamma_4^+$&1&1&-&$\Delta_4$&$\Sigma_1$&-&$\Lambda_3$&-\\
&$\Gamma_5^+$&2&2&-&$\Delta_2$,$\Delta_3$&$\Sigma_3$,$\Sigma_4$&-&$\Lambda_5$&-
\end{tabular}
\caption{Compatibility relations in the intersection between every pair of maximal $\mathbf{k}$-vectors, and the lines connecting them in the space group P$4/ncc$ ($130$). The first column gives the maximal $\mathbf{k}$-vectors. The second column gives, at each $\mathbf{k}$-vector, the distinct irreps of the little group into which the elementary band representation of Table \ref{table:subducedirreps} subduces. The third and fourth columns show the dimension and the multiplicity of the irrep, respectively. The subsequent columns give the compatibility relations between the paths connecting each pair of maximal $\mathbf{k}$-vectors. The symbol "-" indicates that the paths that connect the corresponding $\mathbf{k}$-vectors are redundant as per Sec.~\ref{subsec:independentpaths}, and so can be omitted in the analysis. There are nine pairs of non-null independent paths: the nine bolded paths in Table \ref{table:paths}.}
\label{table:subducedcomprels}
\end{table}

Using this data, we can calculate the possible connectivity graphs that consist of disconnected branches. {We will procede with the following steps:
\begin{enumerate}
	\item First we enumerate all the distinguishable ways the irreps at the first vector $\mathbf{k}^M_1$ can be partitioned into different branches. 

	\item We choose one of the potential decompositions obtained in Step {\bf 1}, and determine the total dimension $d$ of the irreps of the little group $G_{\mathbf{k}^M_1}$ in that branch. We also calculate the subduced representations of the little group of each path involving the point $\mathbf{k}^M_1$. 

	\item We combine the irreps (direct sum of irreps) at the second (in our $(N,\Omega)$ ordering) maximal $\mathbf{k}$ vector $\mathbf{k}^M_2$ in all the possible ways to get branches with dimension $d$. For these possible branches we also calculate the subduced representations of the little group of each path involving $\mathbf{k}^M_2$. 

	\item We compare the sets of irreps along the common path between $\mathbf{k}^M_1$ and $\mathbf{k}^M_2$. If the two sets are identical, then this connection satisfies the compatibility relations and represents part of a valid connectivity graph. If the set of irreps is not the same, then the possible branch is discarded. 

	\item We return iteratively to Step {\bf 1} for each branch that is not discarded in Step {\bf 4}, at each iteration adding the next maximal $\mathbf{k}$-vector $\mathbf{k}^M_i$ in our ordering, and repeating the same calculations for all sets of connected vectors. We compare, by pairs, the sets of irreps along the common paths between the most recently introduced $\mathbf{k}^M_i$ and each of the previously added maximal $\mathbf{k}$ vectors. Finally, we keep all the possible branches that fulfill all the compatibility relations along all the intermediate paths. This procedure either ends in a valid disconnected subgraph of the connectivity graph of the band representation, or terminates before all maximal $\mathbf{k}$-vectors are added to the branch. 
\end{enumerate}

Let us apply this method to our example in space group P$4/ncc$. In our example, there are 8 possible disconnected sets: the four irreps can be divided into 4 branches, $\{(R_1),(R_1),(R_2),(R_2)\}$, into three branches in three different ways, $\{(R_1),(R_1),(R_2,R_2)\}$, $\{(R_1),(R_1,R_2),(R_2)\}$ and $\{(R_1,R_1),(R_2),(R_2)\}$, or into two branches in four different ways, $\{(R_1,R_1),(R_2,R_2)\}$, $\{(R_1,R_2),(R_1,R_2)\}$, $\{(R_1,R_1,R_2),(R_2)\}$ and $\{(R_1),(R_1,R_2,R_2)\}$.} Except for the two decompositions into 2 branches of 2 irreps, $\{(R_1,R_1),(R_2,R_2)\}$ and $\{(R_1,R_2),(R_1,R_2)\}$, all decompositions include a branch with a single copy of either $R_1$ or $R_2$. Let us take a decomposition with a branch that includes $R_1$ at the point $R$, which has dimension 2. We look first along path $W$. Note that the space group of the example is a non-symmorphic group and that the glide reflections $\{m_{1,0,0}|1/2,0,1/2\}$ and $\{m_{0,1,0}|0,1/2,1/2\}$ {in the little group $G_W$. However, as explained in section \ref{subsec:nonsymmorphic}, in this case there is only one set of compatibility relations needed along every line, in particular along $W$}. We have from the compatibility relations that $R_1$ is connected to $W_1$ and $W_4$ 
According to Table~\ref{table:subducedcomprels}, since all irreps at the $X$ point are 2-dimensional, only one of them must be included in the branch of $R_1$. But both $X_1$ and $X_2$ are connected to $W_1$ and $W_3$, but not $W_4$. Therefore, there cannot be a branch that includes at $R$ only the irrep $R_1$ in which the compatibility relations are fulfilled. The same is true for a branch that includes only the irrep $R_2$ at the $R$ point.

Next we check the possible decomposition $\{(R_1,R_1),(R_2,R_2)\}$. In the first branch, the compatibility relations that involve the direct sum $R_1\oplus$$R_1$ with dimension 4 are, $R_1\oplus$$R_1\to$ 2 W$_1\oplus$ 2 W$_4$. At the $X$ point, there is no way to get a direct sum of two irreps whose compatibility relations give the same set as for $R_1\oplus R_1$. Therefore the pair ($R_1,R_1$) cannot be the only two irreps at $R$ of a branch in which the compatibility relations are fulfilled.

Finally, we check the possible decomposition $\{(R_1,R_2),(R_1,R_2)\}$. In this case, there are solutions of the compatibility relations. The Table \ref{table:branch} shows the sets of irreps at each $\mathbf{k}^M$-vec that form a branch that fulfill the compatibility relations. This table can be viewed as a $6\times6$ matrix whose $(i,j)$'th element gives the set of compatibility relations between the direct sum of irreps chosen at the maximal $\mathbf{k}^{M_i}$ point of the reciprocal space and the irreps at the intermediate paths between the $\mathbf{k}^{M_i}$ and $\mathbf{k}^{M_j}$ maximal $\mathbf{k}$-vectors. The subset of irreps chosen at the maximal $\mathbf{k}$-vectors form a branch which fulfills the compatibility relations if this matrix is symmetric. Finally, Table~\ref{table:allbranches} gives all the possible partitions of irreps into two different branches for the elementary band representation of Table~\ref{table:subducedirreps}. This enumerates all disconnected connectivity graphs for this band representation.

\begin{table}[h!]
\begin{tabular}{ll|c|c|c|c|c|c}
&irrep&$R:(0,1/2,1/2)$&$X:(0,1/2,0)$&$M:(1/2,1/2,0)$&$A:(1/2,1/2,1/2)$&$Z:(0,0,1/2)$&$\Gamma:(0,0,0)$\\
\hline
$R:(0,1/2,1/2)$&$R_1\oplus R_2$&-&$W_1$,$W_2$,$W_3$,$W_4$,&-&$T_1$,$T_2$,$T_3$,$T_4$,&4 $U_1$&-\\
\hline
$X:(0,1/2,0)$&$X_1\oplus X_2$&$W_1$,$W_2$,$W_3$,$W_4$,&-&2 $Y_1$&-&-&$\Delta_1$,$\Delta_2$,$\Delta_3$,$\Delta_4$\\
\hline
$M:(1/2,1/2,0)$&$M_1\oplus M_2$&-&2 $Y_1$&-&2 $V_5$&-&$\Sigma_1$,$\Sigma_2$,$\Sigma_3$,$\Sigma_4$\\
\hline
$A:(1/2,1/2,1/2)$&$A_1\oplus A_2$&$T_1$,$T_2$,$T_3$,$T_4$&-&2 $V_5$&-&2 $S_1$&-\\
\hline
$Z:(0,0,1/2)$&$Z_2\oplus Z_3$&4 $U_1$&-&-&2 $S_1$&-&$\Lambda_1$,$\Lambda_4$,$\Lambda_5$\\
\hline
$\Gamma:(0,0,0)$&$\Gamma_1^+\oplus \Gamma_3^+\oplus \Gamma_5^+$&-&$\Delta_1$,$\Delta_2$,$\Delta_3$,$\Delta_4$&$\Sigma_1$,$\Sigma_2$,$\Sigma_3$,$\Sigma_4$&-&$\Lambda_1$,$\Lambda_4$,$\Lambda_5$&-\\
\end{tabular}
\caption{Sets of irreps at the maximal $\mathbf{k}$-vectors that form a branch of the elementary band representation of Table \ref{table:branch}, and the sets of compatibility relations along the paths that join every pair. {Note that the symmetry of this table reflects the fact that the compatibility relations are satisfied.}}
\label{table:branch}
\end{table}

\begin{table}[h!]
\begin{tabular}{ll|l}
solution&branch 1&branch 2\\
\hline
1&$R_1$,$R_2$,$X_1$,$X_2$,$M_1$,$M_2$,$A_1$,$A_2$,$Z_2$,$Z_3$,$\Gamma_1^{+}$,$\Gamma_3^{+}$,$\Gamma_5^{+}$&$R_1$,$R_2$,$X_1$,$X_2$,$M_3$,$M_4$,$A_3$,$A_4$,$Z_1$,$Z_4$,$\Gamma_2^{+}$,$\Gamma_4^{+}$,$\Gamma_5^{+}$\\
2&$R_1$,$R_2$,$X_1$,$X_2$,$M_1$,$M_2$,$A_1$,$A_2$,$Z_2$,$Z_4$,$\Gamma_1^{+}$,$\Gamma_3^{+}$,$\Gamma_5^{+}$&$R_1$,$R_2$,$X_1$,$X_2$,$M_3$,$M_4$,$A_3$,$A_4$,$Z_1$,$Z_3$,$\Gamma_2^{+}$,$\Gamma_4^{+}$,$\Gamma_5^{+}$\\
3&$R_1$,$R_2$,$X_1$,$X_2$,$M_1$,$M_2$,$A_1$,$A_2$,$Z_1$,$Z_3$,$\Gamma_2^{+}$,$\Gamma_4^{+}$,$\Gamma_5^{+}$&$R_1$,$R_2$,$X_1$,$X_2$,$M_3$,$M_4$,$A_3$,$A_4$,$Z_2$,$Z_4$,$\Gamma_1^{+}$,$\Gamma_3^{+}$,$\Gamma_5^{+}$\\
4&$R_1$,$R_2$,$X_1$,$X_2$,$M_1$,$M_2$,$A_1$,$A_2$,$Z_1$,$Z_4$,$\Gamma_2^{+}$,$\Gamma_4^{+}$,$\Gamma_5^{+}$&$R_1$,$R_2$,$X_1$,$X_2$,$M_3$,$M_4$,$A_3$,$A_4$,$Z_2$,$Z_3$,$\Gamma_1^{+}$,$\Gamma_3^{+}$,$\Gamma_5^{+}$\end{tabular}
\caption{All possible decompositions of the elementary band representation of Table \ref{table:subducedirreps} into two branches. Each row represents a different disconnected connectivity graph. The second and third columns give the set of irreps of the little groups of each maximal $\mathbf{k}$-vector in the different connected subgraphs.}
\label{table:allbranches}
\end{table}

\section{Data Records}\label{sec:filedesc}
\begin{figure}[H]
\centering
\includegraphics[width=0.9\textwidth]{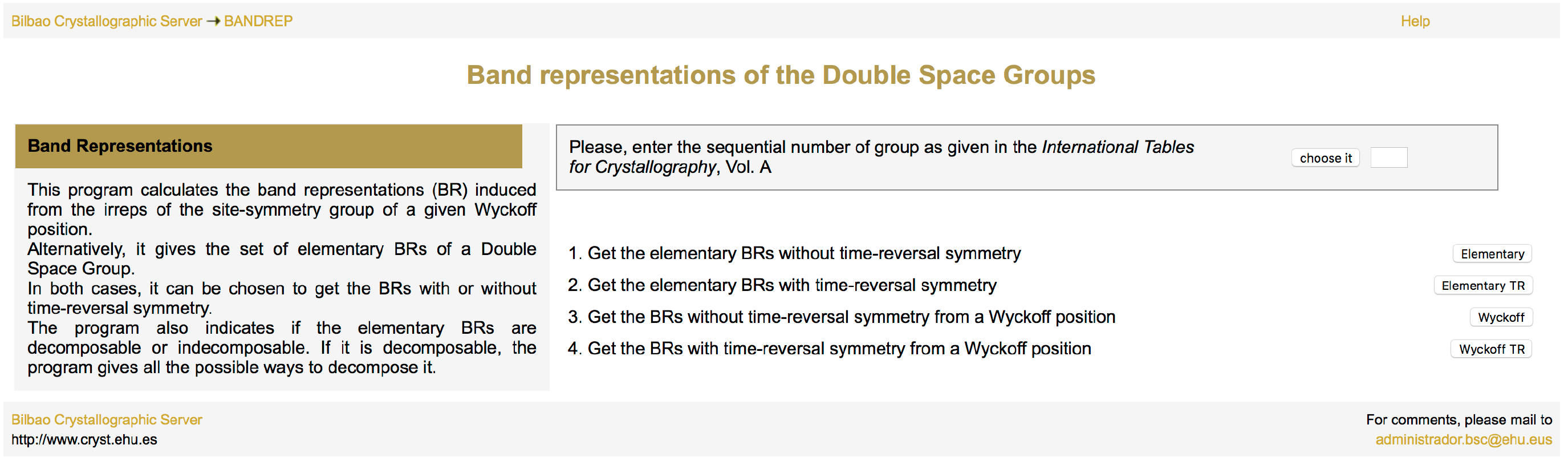}
\caption{Main input screen for the BANDREP program.}\label{data:bandrepinput}
\end{figure}

Using the algorithms described in Sec.~\ref{sec:methods} along with the group-theoretic data computed in Ref.~\onlinecite{grouptheory}, we have computed the minimal paths through the Brillouin zone for each of the $230$ space groups needed to determine the connectivity of energy bands both with and without TR symmetry, as well as all possible disconnected connectivity graphs for all elementary and physically elementary band representations. We have compiled the output of the algorithms described in Sec.~\ref{sec:methods} into different subprograms, all contained within the ``BANDREP'' application on the Bilbao Crystallographic Server (\url{www.cryst.ehu.es/cryst/bandrep}). The main input screen for this application is shown in Fig.~\ref{data:bandrepinput}. Here we will focus only on those features directly related to connectivity graphs.

\begin{figure}[ht]
\centering
\includegraphics[width=\textwidth]{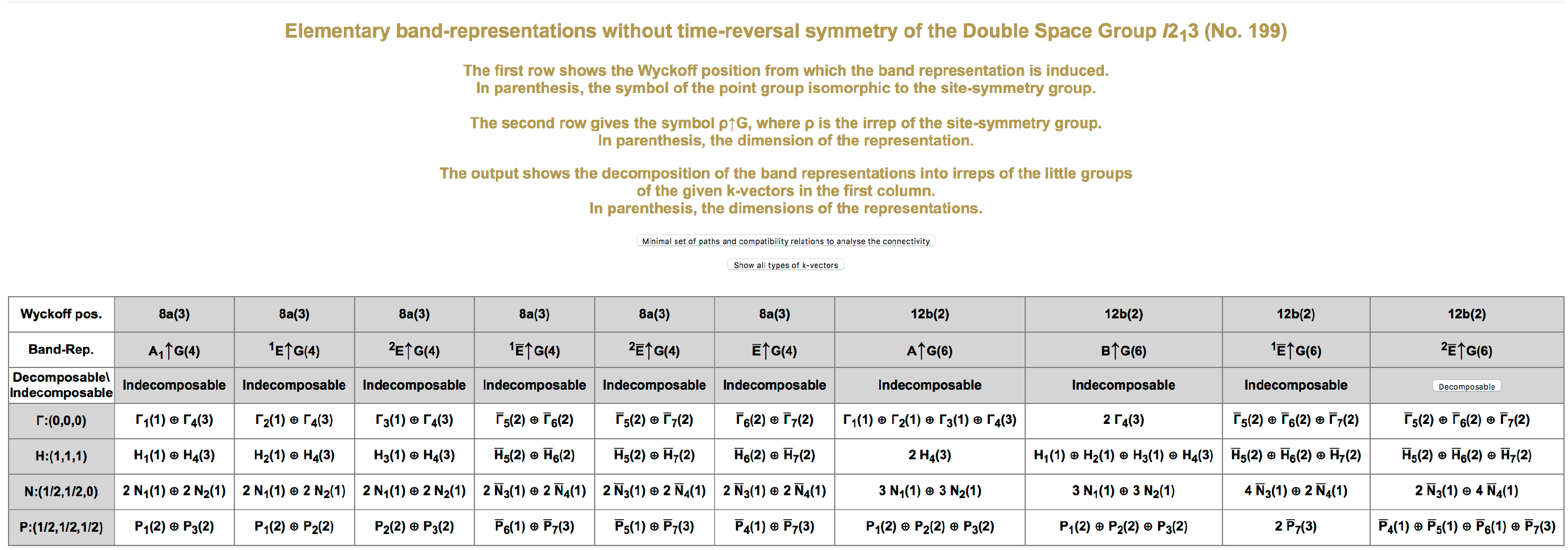}
\caption{Output of BANDREP for the elementary band representations in SG $I2_13$ ($199$) without TR. There is one decomposable elementary band representation. It is induced from the two-dimensional ${}^2\bar{E}$ representation of the site-symmetry group of the $12b$ Wyckoff position.}\label{data:bands_decomposable199}
\end{figure}

\begin{figure}[ht]
\centering
\includegraphics[width=0.5\textwidth]{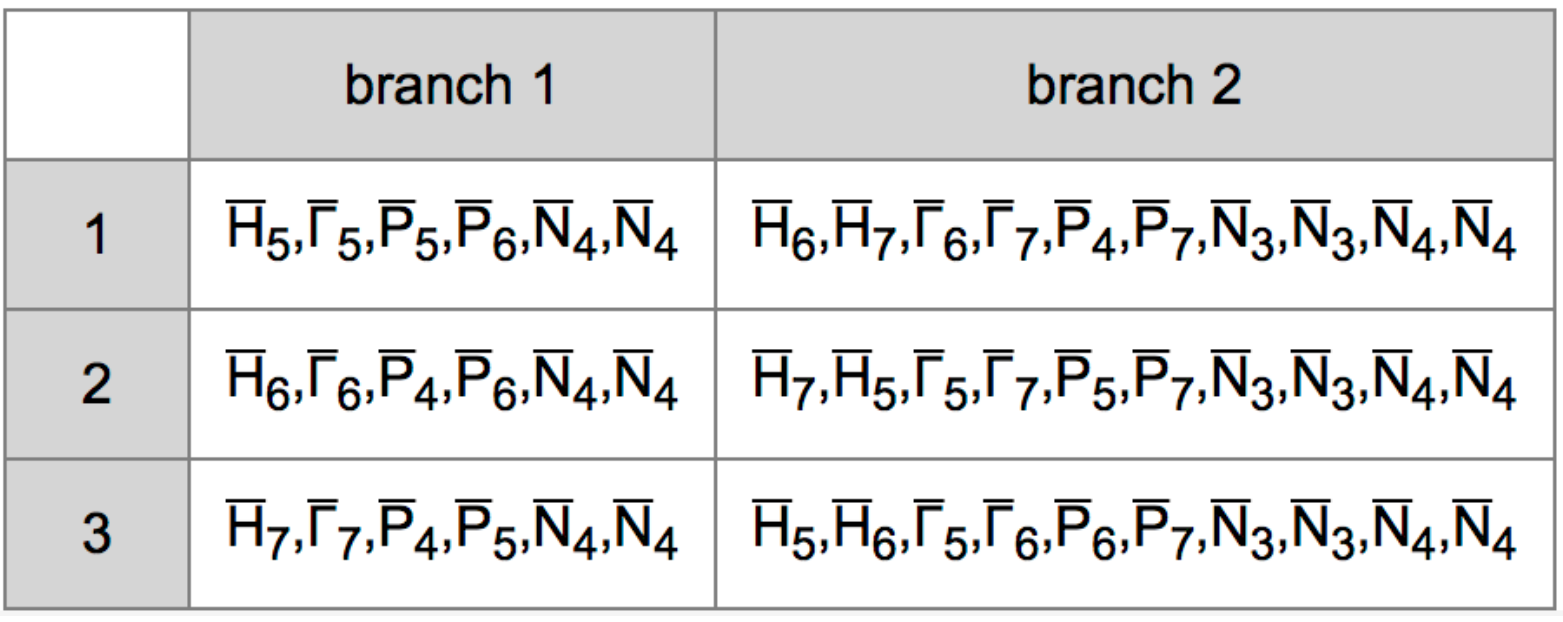}
\caption{Possible decompositions of the elementary band representation in SG $I2_13$ ($199$) induced from the ${}^2\bar{E}$ representation of the site symmetry group of the $12b$ maximal Wyckoff position.}\label{data:bandgraphs199}
\end{figure}

First, let us describe how to access the disconnected solutions for the connectivity graphs of a band representation. Entering the number of a space group, and clicking on either the ``Elementary'' or ``Elementary TR'' buttons gives a table of all elementary or physically elementary band representations in the given space group, respectively. Band representations are listed according to Wyckoff position, and the irreducible representation of the site-symmetry group from which they are induced. In addition to the little group representations subduced at each maximal $\mathbf{k}$-vector, for each band representation the output table contains a row labelled ``Decomposable\textbackslash Indecomposable,'' which indicates whether or not a disconnected connectivity graph exists for the given band representation. In Fig.~\ref{data:bands_decomposable199}, we show the output of selecting ``Elementary'' for the space group $I2_13$ ($199$). In particular, there is one decomposable elementary band representation. It is induced from the ${}^2\bar{E}$ ($\bar{\Gamma}_3$) representation of the site-symmetry group of the $12b$ Wyckoff position, which is isomorphic to the point group $C_2$. For this band representation -- and more generally for any band representation with disconnected connectivity graphs -- the entry in the ``Decomposable\textbackslash Indecomposable'' row is a clickable button. The output of clicking this button is a list of all possible ways of partitioning connectivity graphs into disconnected components. This data is given in the format of Sec.~\ref{sec:luisgraph} and Table~\ref{table:allbranches}; each row corresponds to a different disconnected solution to the compatibility relations, and each column gives the little group representations subduced at each maximal $\mathbf{k}$-vector in each branch (disconnected component). Fig.~\ref{data:bandgraphs199} shows this output for the decomposable band representation ${}^2\bar{E}\uparrow G$ induced from the $12b$ position in SG $I2_13$ ($199$). We see that there are three possible disconnected connectivity graphs, each with two disconnected components.

\begin{figure}[ht]
\centering
\includegraphics[width=\textwidth]{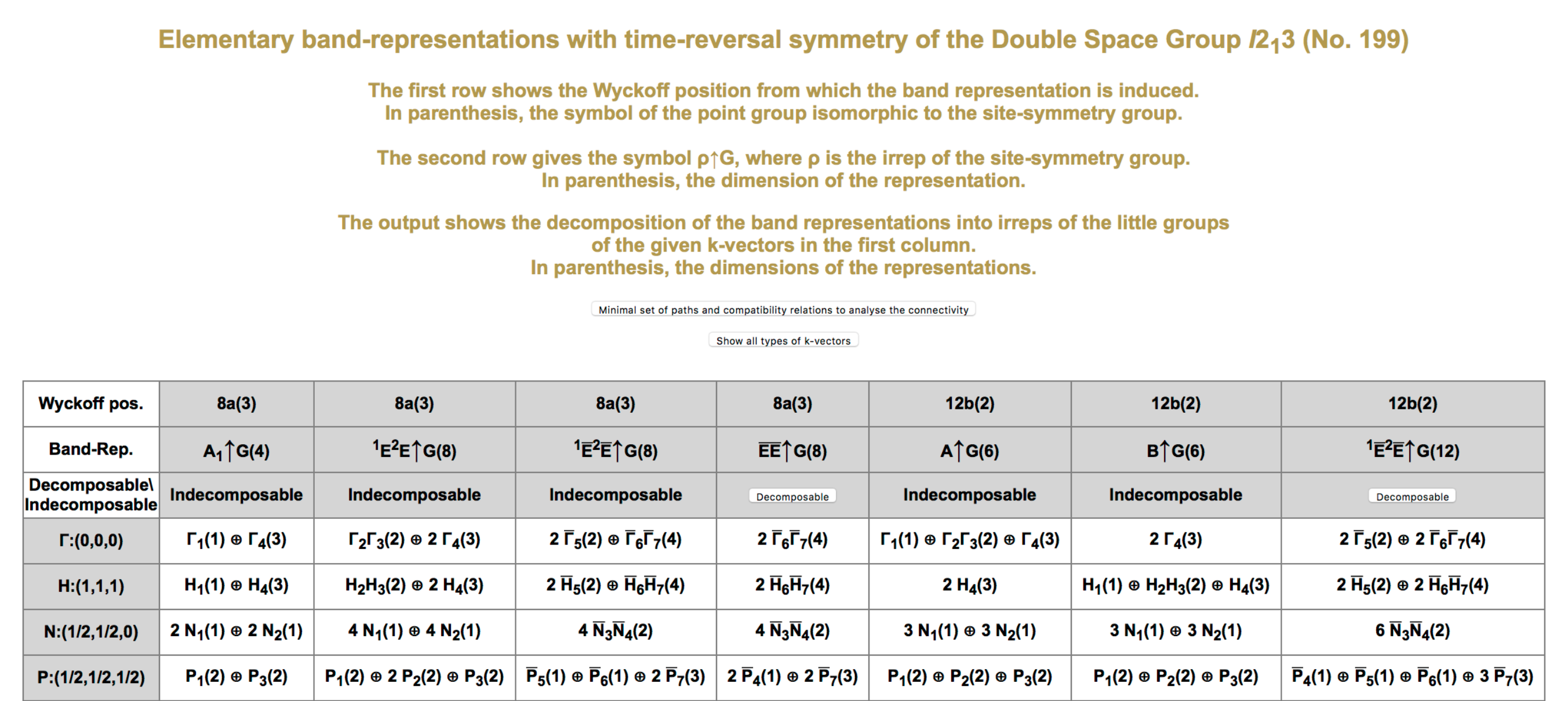}
\caption{Output of BANDREP for the physically elementary band representations in SG $I2_13$ ($199$). There are two decomposable physically elementary band representations, induced from the $8a$ and $12b$ maximal Wyckoff position}\label{data:bands_decomposableTR199}
\end{figure}

\begin{figure}[ht]
\centering
\includegraphics[width=0.6\textwidth]{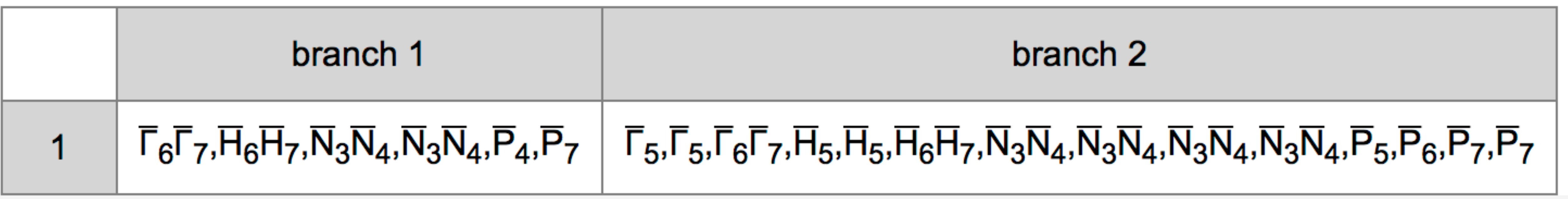}
\caption{Decomposition of the elementary band representation in SG $I2_13$ ($199$) induced from the ${}^1\bar{E{}^2\bar{E}}$ physically irreducible representation of the site symmetry group of the $12b$ maximal Wyckoff position.}\label{data:bandgraphsTR199}
\end{figure}

To obtain the analogous information for the physically elementary band representations with TR symmetry, we can click instead the ``Elementary TR'' button on the main input screen. This output for space group $I2_13$ ($199$) is shown in Fig.~\ref{data:bands_decomposableTR199}. We see that with TR symmetry, there are now two decomposable physically elementary band representations. The first is induced from the physically irreducible $\bar{E}\bar{E}$ ($\bar{\Gamma}_4\bar{\Gamma}_4$) representation of the site-symmetry group of the $8a$ position, isomorphic to the point group $C_3$. The second decomposable physically elementary band representation is induced from the ${}^1\bar{E}{}^2\bar{E}$ ($\bar{\Gamma}_3\bar{\Gamma}_4$) representation of the site-symmmetry group of the $12b$ Wyckoff position, which is isomorphic to the point group $C_2$ . In Fig.~\ref{data:bandgraphsTR199} we show the possible disconnected connectivity graphs for this latter band representation. It turns out that in this case there is only one allowed disconnected connectivity graph, with two branches.

In addition to the connectivity graphs, we also give, for each space group, the minimal list of paths through the BZ and the associated compatibility relations needed to construct the full connectivity graphs from the little group representations at the maximal $\mathbf{k}$-vectors. From the table of band representations accessed from either the ``Elementary'' or ``Elementary TR'' function, this data can be accessed by clicking the button labelled ``Minimal set of paths and compatibility relations to analyse the connectivity.'' The location of this button above the table of band representations can be seen in Figs.~\ref{data:bands_decomposable199} and \ref{data:bands_decomposableTR199}. The output of this application gives two tables. The first table lists the minimal set of connections between maximal $\mathbf{k}$-vectors, given in the format of Table~\ref{table:paths}. It has three columns: each row gives two maximal $\mathbf{k}$-vectors in the first and third column which are connected by the non-maximal $\mathbf{k}$-vector in the second column.

Directly below the table of $\mathbf{k}$-vectors, we display the compatibility relations along each of the listed connections. This table has five columns. The first, third, and fifth columns correspond to the first maximal, intermediate, and second maximal $\mathbf{k}$-vector columns given in the table of connections, while the second and fourth columns give the compatibility relations along each connection. For each little group representation of the maximal $\mathbf{k}$-vectors, the compatibility relations are given in the format of Eq.~(\ref{eq:comprelformat}). For those non-symmorphic groups that require two different sets of compatibility relations related by monodromy, the second set is given immediately next to the first.

\begin{figure}[H]
\centering
\includegraphics[width=\textwidth]{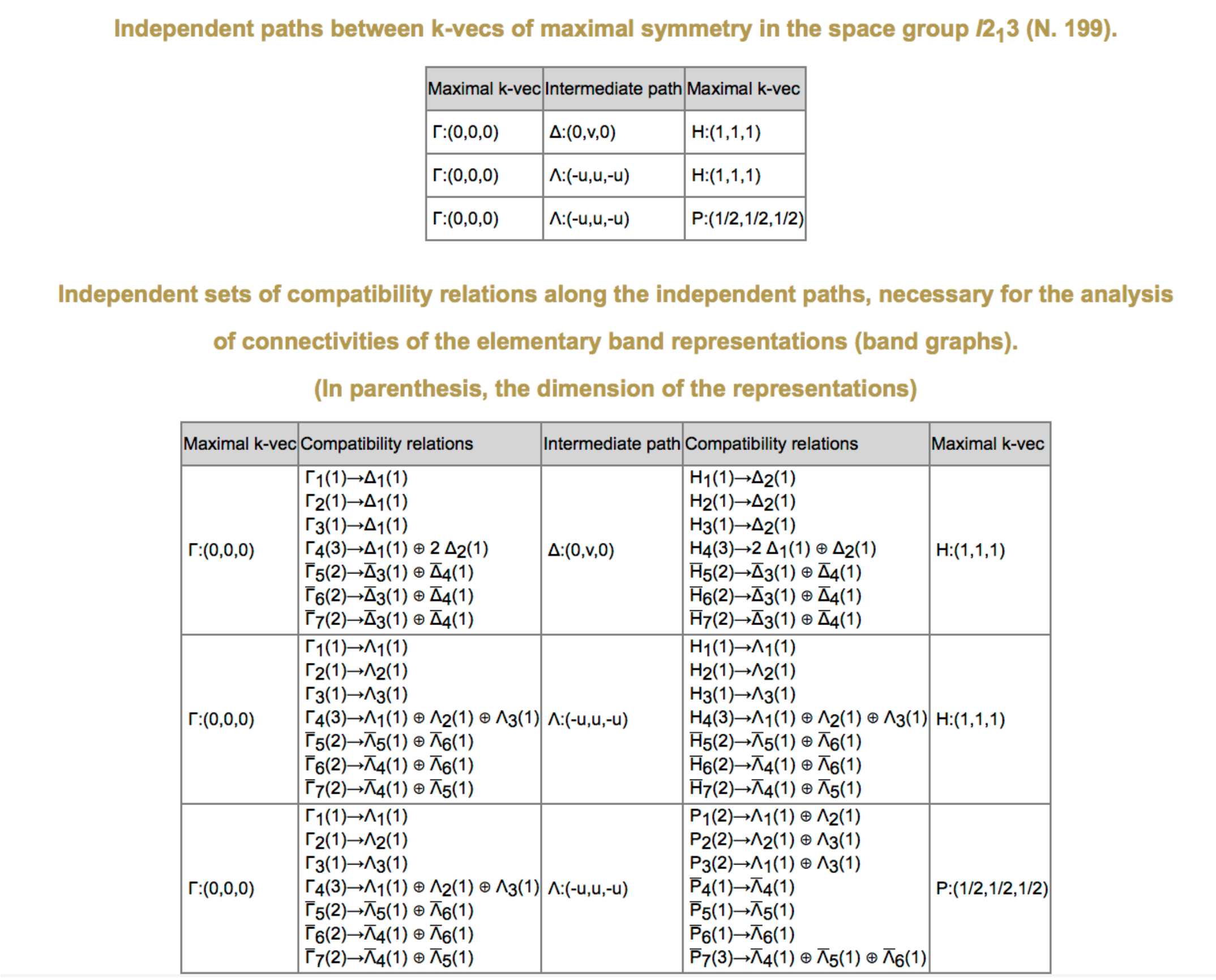}
\caption{Minimal path and associated compatibility tables for SG $I2_13$ ($199$) without TR symmetry.}\label{data:comprel199}
\end{figure}

As an example, we show in Fig.~\ref{data:comprel199} the set of paths and compatibility relations for SG $I2_13$ ($199$) without TR symmetry, obtained by clicking the ``Minimal set of paths and compatibility relations to analyse the connectivity.'' button in Fig.~\ref{data:bands_decomposable199}. We see that there are only three maximal $\mathbf{k}$-vectors that determine the connectivity, $\Gamma$, $H$, and $P$. There are three essential connections,
\begin{eqnarray}\label{eq:con199}
\Gamma&\leftrightarrow&\Delta\leftrightarrow H,\\
\Gamma&\leftrightarrow&\Lambda\leftrightarrow H,\\
\Gamma&\leftrightarrow&\Lambda\leftrightarrow P.
\end{eqnarray}
Although this group is non-symmorphic, we see from the compatibility table that only one set of compatibility relations is needed along each connection. This is due to the additional constraints imposed by the cubic threefold rotation.

\begin{figure}[H]
\centering
\includegraphics[width=\textwidth]{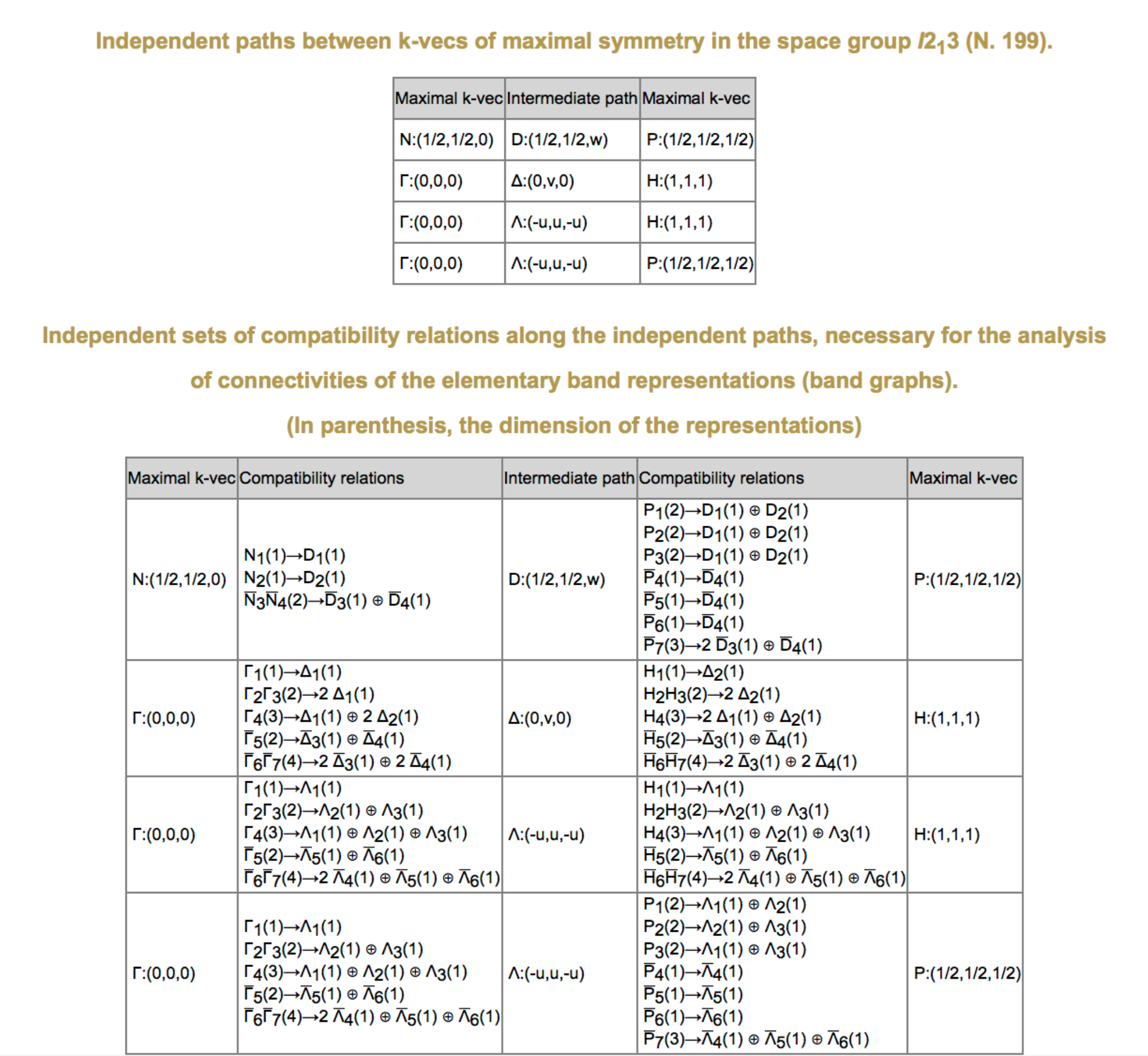}
\caption{Minimal path and associated compatibility tables for SG $I2_13$ ($199$) in the presence of TR symmetry.}\label{data:comprelTR199}
\end{figure}

Clicking on the analogous button in the output of Fig.~\ref{data:bands_decomposableTR199} gives the minimal paths and compatibility relations for this same space group once time-reversal symmetry is included. We show these in Fig.~\ref{data:comprelTR199}. We see immediately that TR singles out an additional (TR-invariant) maximal $\mathbf{k}$-vector, labelled $N$. In addition to the connections in Eq.~(\ref{eq:con199}), we see that with TR we must also consider compatibility along the connection
\begin{equation}
N\leftrightarrow D\leftrightarrow P.
\end{equation} 
Once again, we see from the compatibility table that only one set of compatibility relations is needed for every connection in this space group with TR symmetry.

\section{Technical Validation}\label{sec:valid}

Now that we have produced the data and the applications with which to access it, we will show here an example of how they may be used. We examine the case of graphene {on a graphite (or another symmetry-preserving, lattice-matched substrate which breaks only inversion symmetry)}, corresponding to the Kane-Mele model with inversion-symmetry breaking. This is described by the three-dimensional space group $P6mm$ ($183$). We will see how we can recover the full topological phase diagram using the graph output files, and insodoing give a consistency check on our data. 

\begin{figure}[H]
\centering
\includegraphics[width=0.9\textwidth]{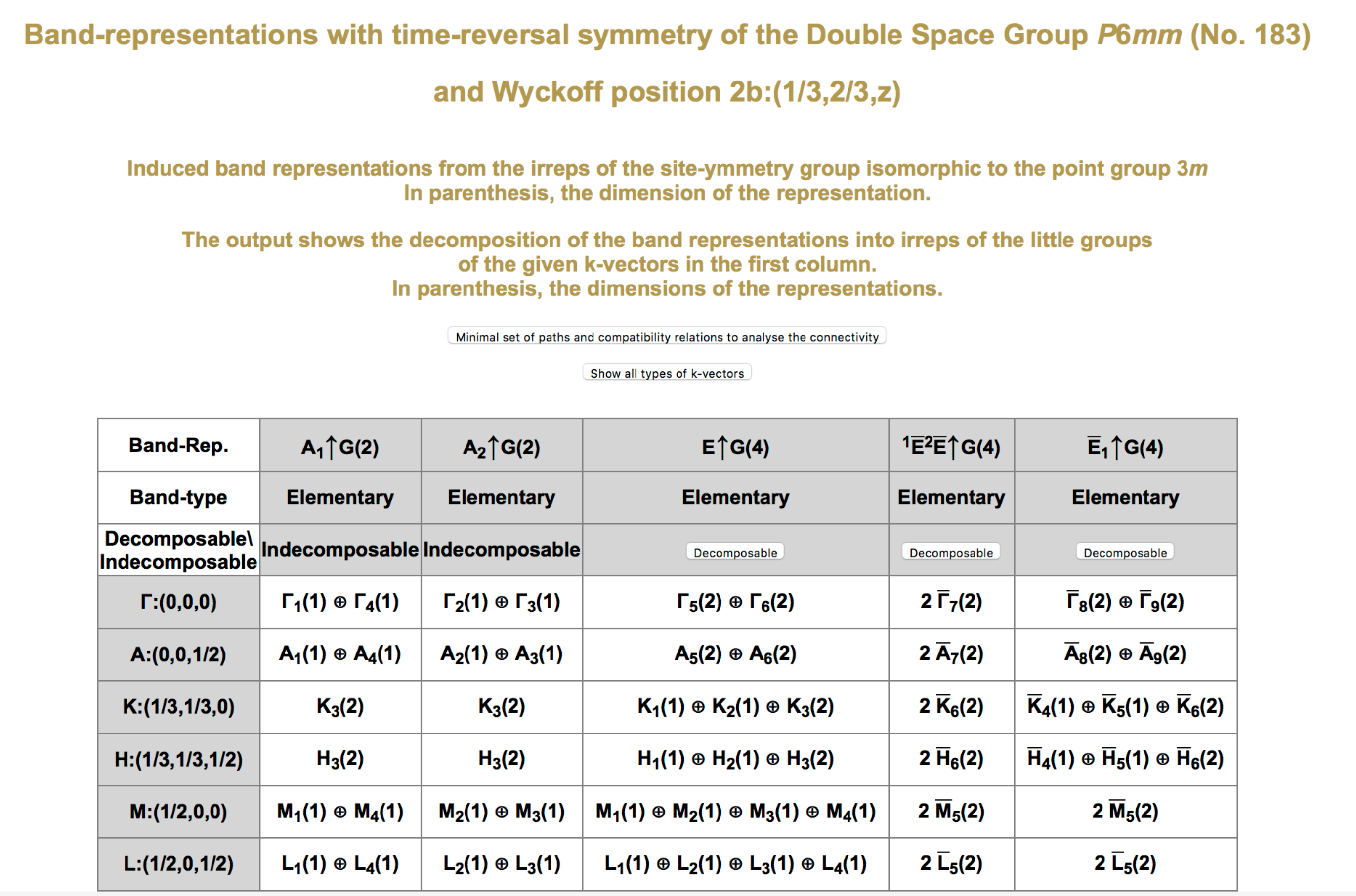}
\caption{Output of BANDREP for the $2b$ position of SG $P6mm$ ($183$) with TR symmetry.}\label{fig:1832breps}
\end{figure}

The relation between the topological phases of graphene and the connectivity of elementary band representations was computed first in Refs.~\onlinecite{NaturePaper,EBRtheory}. Here we will show how to recover these computations using the applications we have produced. The carbon atoms in graphene sit at the $2b$ Wyckoff position of space group $P6mm$ ($183$). The site-symmetry group of this position is isomorphic to the point group $C_{3v}$ ($3m$), generated by a threefold rotation $C_{3z}$ about the $z$-axis (normal to the plane) and the vertical mirror $m_y$. By consulting the data presented in Refs.~\onlinecite{NaturePaper,grouptheory}, we can see that spinful $p_z$ orbitals transform in the two-dimensional $\bar{\Gamma}_6$ representation of this group. Next, we consult the BANDREP program for this space group. Since we are interested in orbitals at the $2b$ position only with time-reversal symmetry, we may use the ``Wyckoff TR'' option\cite{grouptheory}. This outputs the physically elementary band representations induced only from the $2b$ Wyckoff position; the output is shown in Fig.~\ref{fig:1832breps}.

\begin{figure}[H]
\centering
\includegraphics[width=0.6\textwidth]{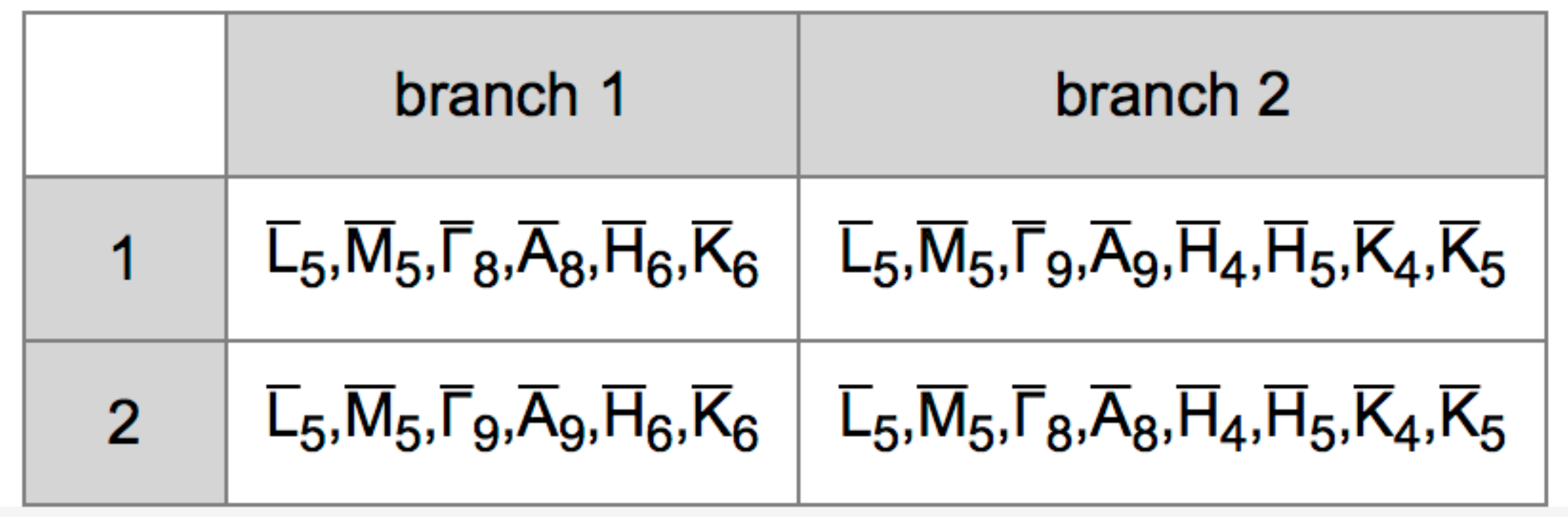}
\caption{Possible decompositions of the physically elementary band representation induced from the $\bar{E}_1\uparrow G$ representation of the site-symmetry group of the $2b$ position in SG $P6mm$ ($183$).}\label{fig:183graphs}
\end{figure}

From this we see that the band representation induced from the $\bar{E}_1$ $(\bar{\Gamma}_6)$ representation at the $2b$ site is decomposable. From the first column of the table, we see that the maximal $\mathbf{k}$ vectors in this space group are labelled $\Gamma,K,M,A,H,L$. Since we are only interested in the two-dimensional system, we need only concern ourselves with the subset $\Gamma,K,M$ of maximal $\mathbf{k}$-vectors at $k_z=0$. Furthermore, the little group of all vertical lines in this space group is the same as the unitary subgroup of the little group of each endpoint, and so the compatibility relations along the vertical are trivial. Thus, we can find all disconnected compatibility graphs for the $2D$ system from the BANDREP application by simply looking at the $k_z=0$ subgraph of every $3D$ connectivity graph. Knowing this, we examine the output of the ``Decomposable'' option for the $\bar{E}_1\uparrow G$ band representation induced from the $2b$ position of SG $P6mm$, given in Fig.~\ref{fig:183graphs}
We see that there are two different disconnected connectivity graphs. In the first, the $2D$ system has one connected component containing the $\bar{\Gamma}_8,\bar{K}_6,$ and $\bar{M}_5$ little group representations, while the other component contains $\bar{\Gamma}_9,\bar{K}_4,\bar{K}_5$ and $\bar{M}_5$. The second disconnected solution has $\bar{\Gamma}_8$ and $\bar{\Gamma}_9$ interchanged. This matches precisely the result of Ref.~\onlinecite{NaturePaper} obtained by a direct analysis of the compatibility relations. These disconnected band graphs correspond to the topologically disconnected bands of the Kane-Mele model of graphene with Rashba spin-orbit coupling.

\section{Usage Notes}\label{sec:notes}

All of the data and applications described in Sec.~\ref{sec:filedesc} can be accessed via the ``BANDREP'' program at the Bilbao Crstallographic Server, accessible at \url{http://www.cryst.ehu.es/cryst/bandrep}. In conjunction with the group theory applications described in Ref.~\onlinecite{grouptheory}, and hosted on the Bilbao Crystallographic Server, all information on type $(1,1)$ topological phases\cite{NaturePaper} may be deduced. Additionally, the algorithms described in Sec.~\ref{sec:methods} may be used along with any list of little group representations to deduce the existence of type $(1,2)$ and type $(2,2)$ band-inversion topological insulators.

\begin{acknowledgments}
BB would like to thank Ida Momennejad and Dustin Ngo for fruitful discussions. MGV would like to thank Gonzalo Lopez-Garmendia for help with computational work. BB, JC, ZW, and BAB acknowledge the hospitality of the Donostia International Physics Center, where parts of this work were carried out. JC also acknowledges the hospitality of the Kavli Institute for Theoretical Physics, and BAB also acknowledges the hospitality and support of the \'{E}cole Normale Sup\'{e}rieure and Laboratoire de Physique Th\'{e}orique et Hautes Energies. The work of MVG was supported by FIS2016-75862-P and FIS2013-48286-C2-1-P national projects of the Spanish MINECO. The work of LE and MIA was supported by the Government of the Basque Country (project IT779-13)  and the Spanish Ministry of Economy and Competitiveness and FEDER funds (project MAT2015-66441-P). ZW and BAB, as well as part of the development of the initial theory and further ab-initio work, were supported by the NSF EAGER Grant No. DMR-1643312, ONR - N00014-14-1-0330, ARO MURI W911NF-12-1-0461, and NSF-MRSEC DMR-1420541. The development of the practical part of the theory, tables, some of the code development, and ab-initio work was funded by Department of Energy de-sc0016239, Simons Investigator Award, the Packard Foundation, and the Schmidt Fund for Innovative Research. 

\end{acknowledgments}

\bibliography{References_Luis}
\end{document}